\documentclass[acmsmall,authorversion,nonacm,screen]{acmart}

\usepackage{booktabs}   
\usepackage{listings}
\usepackage{caption, subcaption} 
\usepackage{enumitem}
\usepackage{todonotes}
\usepackage{physics}
\usepackage{bm}
\usepackage{wasysym}
\usepackage[normalem]{ulem}
\usepackage{algorithmicx}
\usepackage{algorithm}
\usepackage{algpseudocode}
\usepackage{smartref}
\usepackage{multicol}
\usepackage[frozencache,cachedir=.]{minted}

\newminted[mlir]{text}{fontsize=\footnotesize, escapeinside=||, mathescape}
\newmintinline[mlirinline]{text}{}

\newminted[mlirln]{text}{fontsize=\footnotesize, escapeinside=||, mathescape,linenos}

\lstdefinestyle{mystyle}{
    basicstyle=\ttfamily\footnotesize,
    breaklines=true,
}

\graphicspath{{figures/barvinok/}{figures/barvinok_indexing}}
\begin{document}

\newcommand{\RA}[1]{{#1}}
\newcommand{\RB}[1]{{#1}}
\newcommand{\RC}[1]{{#1}}
\newcommand{\RAll}[1]{{#1}}
\newcommand{\Rev}[1]{{\color[HTML]{000000}{#1}}}

\newcommand{\Hlgray}{\makebox[0pt][l]{\color[HTML]{DDDDDD}\rule[-4pt]{0.8\linewidth}{10pt}}}
\newcommand{\Hgray}{\makebox[0pt][l]{\color[HTML]{BBBBBB}\rule[-4pt]{0.8\linewidth}{10pt}}}

\newcommand{\amirsh}[1]{{\color{red}Amir: #1}}
\newcommand{\note}[1]{{\color{blue}Note: #1}}
\newcommand{\mh}[1]{{\color{orange}Mathieu: #1}}
\newcommand{\ebauer}[1]{{\color{violet}Emilien: #1}}
\newcommand{\mahdi}[1]{{\color{blue}Mahdi: #1}}

\newcommand{\systemname}{{\sc DASTAC}}
\newcommand{\structtensor}{{\sc StructTensor}}
\newcommand{\stur}{{\sc STUR}}
\newcommand{\smartpara}[1]{\noindent \textbf{#1.}}
\newcommand{\lemexp}[1]{\noindent \normalfont {#1.}}
\newcommand{\translatebegin}{$\llbracket$}
\newcommand{\translateend}{$\rrbracket$}
\newcommand{\translate}[1]{\translatebegin{}\text{#1}\translateend{}}
\newcommand{\metavar}[1]{$#1$}
\newcommand{\metavars}[1]{$\MakeLowercase{#1}$}
\newcommand{\binop}{\diamond}
\newcommand{\domain}[1]{FV(#1)}

\newcommand{\matrixmult}{\cdot}
\newcommand{\tab}{$~~~$}
\newcommand{\argvec}[1]{\bm{#1}}
\newcommand\sbullet[1][.75]{\mathbin{\vcenter{\hbox{\scalebox{#1}{$\bullet$}}}}}
\newcommand{\vectorize}[1]{vec(#1)}
\newcommand{\hcat}{||}
\newcommand{\vcat}{\RA{/\!/}}

\newcommand{\TT}{\mathcal{T}}
\newcommand{\RR}{\mathbb{R}}
\newcommand{\II}{\mathbb{I}}
\newcommand{\BB}{\mathbb{B}}
\newcommand{\forget}[1]{\lfloor #1 \rfloor}

\definecolor{colg}{rgb}{0.1,0.7,0.1}
\definecolor{colr}{rgb}{0.7,0.1,0.1}
\definecolor{colb}{rgb}{0.1,0.1,0.7}
\definecolor{colbb}{rgb}{0.1,0.1,0.6}
\newcommand{\supfull}{{\color{colg}{\CIRCLE}}}
\newcommand{\suphalf}{{\color{colb}{\LEFTcircle}}}
\newcommand{\supnone}{{\color{colr}{\Circle}}}

\newcommand{\examplepara}[1]{\smartpara{Example - #1}}

\definecolor{editcol}{rgb}{0.8, 0.3, 0.7}
\newcommand{\edit}[1]{{\color{editcol}#1}}
\newcommand{\remove}[1]{{\color{colr}\sout{#1}}}
\newcommand{\rewrite}[2]{{\edit{#1} \remove{#2}}}


\title{Compressing Structured Tensor Algebra}

\author{Mahdi Ghorbani$^*$}
\orcid{0000-0003-3686-2158}
\email{Mahdi.Ghorbani@ed.ac.uk}
\affiliation{%
  \institution{University of Edinburgh}
  \city{Edinburgh}
  \country{UK}
}

\author{Emilien Bauer$^*$}
\orcid{}
\email{Emilien.Bauer@ed.ac.uk}
\affiliation{%
  \institution{University of Edinburgh}
  \city{Edinburgh}
  \country{UK}
}

\author{Tobias Grosser}
\orcid{}
\email{tobias.grosser@cst.cam.ac.uk}
\affiliation{%
  \institution{University of Cambridge}
  \city{Cambridge}
  \country{UK}
}

\author{Amir Shaikhha}
\orcid{0000-0002-9062-759X}
\email{amir.shaikhha@ed.ac.uk}
\affiliation{%
  \institution{University of Edinburgh}
  \city{Edinburgh}
  \country{UK}
}

\renewcommand{\shortauthors}{Mahdi Ghorbani, Emilien Bauer, Tobias Grosser, Amir Shaikhha}

\begin{abstract}
Tensor algebra is a crucial component for data-intensive workloads such as machine learning and scientific computing. As the complexity of data grows, scientists often encounter a dilemma between the highly specialized dense tensor algebra and efficient structure-aware algorithms provided by sparse tensor algebra. In this paper, we introduce DASTAC, a framework to propagate the tensors's captured high-level structure down to low-level code generation by incorporating techniques such as automatic data layout compression, polyhedral analysis, and affine code generation. Our methodology reduces memory footprint by automatically detecting the best data layout, heavily benefits from polyhedral optimizations, leverages further optimizations, and enables parallelization through MLIR. Through extensive experimentation, we show that DASTAC achieves 1 to 2 orders of magnitude speedup over TACO, a state-of-the-art sparse tensor compiler, and StructTensor, a state-of-the-art structured tensor algebra compiler, with a significantly lower memory footprint.

\end{abstract}


\maketitle
\def\thefootnote{*}\footnotetext{These authors contributed equally to this work}\def\thefootnote{\arabic{footnote}}

\section{Introduction}
\label{sec:introduction}

\begin{figure}
    \centering
    \includegraphics[height=3cm]{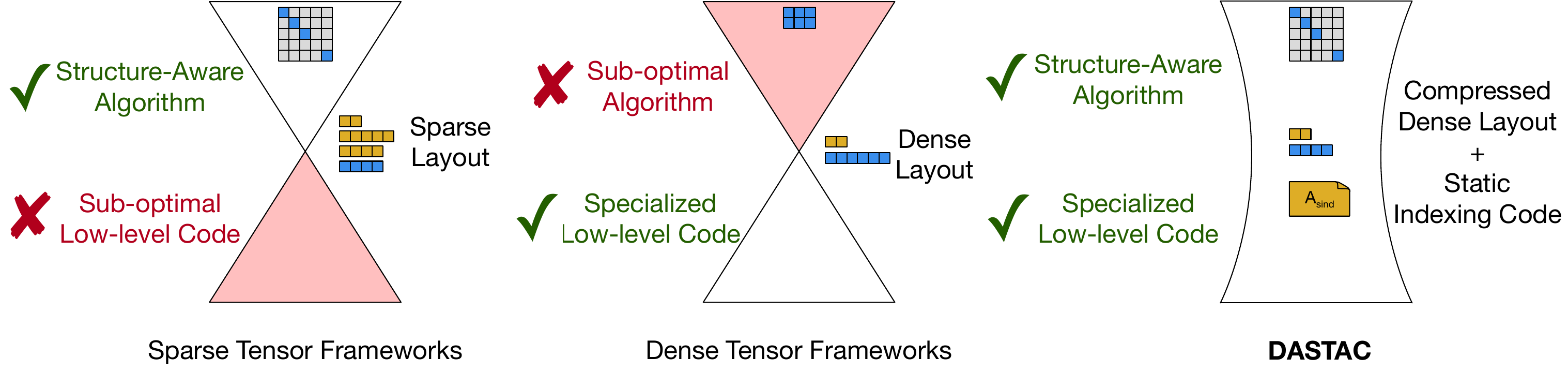}
    \caption{Comparison of tensor processing frameworks. \systemname{} is the first code generation framework that combines the algorithmic optimizations known from sparse tensor algebras with performance-optimized low-level code known from decades of tuning dense tensors.}
    \label{fig:intro}
    \vspace{-0.4cm}
\end{figure}




Tensor algebra is a fundamental component in several data-intensive workloads, such as signal processing~\cite{de1998matrix,nion2010tensor,muti2007survey}, machine learning~\cite{sidiropoulos2017tensor}, computer vision~\cite{panagakis2021tensor}, quantum chemistery~\cite{khoromskaia2018tensor,mutlu2019toward}, and bioinformatics~\cite{teran2019tensor}. Numerous lines of research have attempted to specialize and enhance the performance of the tensor algebra computations either through hardware level (e.g., tensor accelerators~\cite{hegde2019extensor,jia2021tensorlib,gondimalla2019sparten} and TPUs~\cite{jouppi2023tpu}), or software level (e.g., tuned tensor kernels~\cite{wang2014intel,10.1145/77626.79170} and compilation based optimizations~\cite{gareev2018high}).

As the complexity of the data grows, scientists often face a trade-off between highly tuned and specialized frameworks provided for dense tensor algebra computation and efficient and flexible algorithms provided for sparse tensor algebra that leverage the \Rev{sparsity} of input tensors. Dense tensor algebra frameworks~\cite{baghdadi2019tiramisu,TensorComprehension,polly,polyblocks} support a rich set of compile-time optimizations such as vectorization, tiling, and parallelization. This is thanks to well-known memory access patterns and efficient utilization of contiguous memory, despite not knowing the actual data. 

\Rev{Computations over structured tensor algebras are at the core of several important workloads. The structure is an expression of static information on the sparsity and redundancy of the tensors. Sparsity information defines the symbolic set of indices of non-zero values (e.g., diagonal) while redundancy defines a symbolic mapping from the set of indices of repetitive elements to their unique equivalent (e.g., symmetric). Structured matrices have many applications in machine learning~\cite{KhamisNNOS20} and deep learning~\cite{KisselD23}. Diagonal structure used in the ACDC layer~\cite{MoczulskiDAF15}, tridiagonal weight matrices mentioned by~\cite{DumitrasK00}, and structured neural network pruning for large language models~\cite{WangWL20} are examples of applications of matrices with simple structures. Furthermore, more intricate structures such as hyper triangular for covariance matrices~\cite{structtensor} and the butterfly structure used for neural network training~\cite{ChenDLY0RR22} are examples of structured tensor applications.}

Sparse tensor algebra frameworks~\cite{structtensor,chou2022compilation,virtanen2020scipy,spampinato2016basic} provide asymptotically improved algorithms by leveraging the sparsity structure of the data with efficient memory storage requirements and better scalability. However, unlike dense counterparts, sparse frameworks cannot leverage the full computation power of the hardware. This is due to irregular data structures, leading to less predictable memory access patterns and making it challenging to fully leverage the same level of optimizations as their dense counterparts.



In this paper, we introduce \systemname{}, a framework that propagates the high-level structural information of tensor computations down to the low-level highly tuned code generation (cf. Figure~\ref{fig:intro}). \systemname{} leverages a novel symbolic indexing method that compresses an input structured tensor into a densely packed vector representation. In contrast to well-established sparse data layouts like CSR, CSC, or COO, which involve indirect accesses and incur index storage overhead, our method employs efficient direct symbolic index computations.
This direct indexing not only diminishes the overall memory footprint but also opens up opportunities for vectorization. These advantages become more important in scenarios where memory bandwidth is the main bottleneck in program efficiency, as is the case on many widespread contemporary hardware, such as CPUs and GPUs.

\systemname{} achieves the densely assembled data layout by relying on a well-established mathematical foundation: the polyhedral model. First, \systemname{} uses \structtensor{} to capture and propagate the structure throughout the tensor algebra computation and then proceeds to apply polyhedral set and map optimizations over the captured structure. Second, \systemname{} proceeds to find a densely packed data layout for the tensors involved in the computation and compress them automatically by a novel symbolic indexing algorithm, leading to better cache locality and less memory footprint. Finally, \systemname{} generates structure-aware low-level code \Rev{for CPU} using the affine dialect of MLIR~\cite{mlir}, \Rev{enabling effective optimizations at different levels of abstraction.}
Consequently, \systemname{} brings out the best of both dense and sparse tensor algebra worlds together by relying on polyhedral analysis, memory, and complexity efficacy of sparse tensor algebra alongside specialization and compile-time optimization of dense tensor algebra based on polyhedral and compiler optimizations.

Specifically, we make the following contributions:
\begin{itemize}[leftmargin=*]
	\item We introduce \systemname{}, the first framework that integrates the best of both sparse and dense tensor algebra worlds (Section~\ref{sec:overview}). The algorithmic improvements are achieved by leveraging the structure of tensors and this structure is progressively lowered down to the machine-level code. 
	\item \systemname{} introduces a densely assembled data layout that enables low-level specialization opportunities (Section~\ref{sec:datalayout}). The central technique is a novel static indexing algorithm that maps the indices of the input tensor to a contiguous memory buffer.
	\item We provide a progressive code generation algorithm (Section~\ref{sec:compile:codegen}). First, the tensor computations are translated to affine operations provided out-of-the-box by the MLIR framework. Subsequently, at the next stages, the MLIR framework allows \systemname{} to benefit from optimizations such as code motion, common subexpression elimination and parallelization.
	\item We experimentally evaluate \systemname{} against state-of-the-art tensor algebra frameworks and show that leveraging structure while benefiting from our proposed compressed data layout and polyhedral optimizations leads to better performance both sequentially and on multi-threaded scenarios (Section~\ref{sec:experiment}). We show that \systemname{} \Rev{achieves 1 to 2 orders of magnitude speedup over the TACO sparse tensor compiler and StructTensor, a state-of-the-art structured tensor algebra compiler, with a significantly lower memory footprint. We also show that frameworks such as Polygeist cannot recover the optimizations offered by \systemname{}.}
\end{itemize}

\section{Background on Structured Tensor Algebra}
\label{sec:background}

Matrices and tensors can have several structures, such as diagonal, tridiagonal, symmetric, and triangular. Leveraging such structures can improve the computational cost of tensor operations by orders of magnitude. For example, leveraging the matrix structure in sparse matrix-vector multiplication (SpMV), where the matrix is diagonal, can lower the computational complexity from $O(n^2)$ to $O(n)$.

Tensor structures can be classified into two categories: 1) sparsity patterns and 2) redundancy patterns. The sparsity patterns refer to structures such as diagonals where zero and non-zero values are distinguished. Structures with redundancy pattern refers to structures such as symmetry, with unique and repetitive elements. The tensor can be reconstructed using the unique values and a mapping from the indices of redundant elements to the unique ones. 


\stur{}~\cite{structtensor} \Rev{is the \structtensor{}’s unified intermediate representation used for representing both arithmetic and structural information. \structtensor{} relies on structure inference for tensor computations in the form of unique sets, redundancy maps, and compressed tensors. We explain each component of \stur{} and how it works using a running example.}

\Rev{\smartpara{Unique set}}
It represents the \Rev{non-zero and unique (non-repetitive)} values of the input tensor and corresponds to capturing the sparsity pattern. \Rev{For example, the unique set of an $n \times n$ diagonal matrix $M$ represented as $M_U$ is as follows:}

\begin{tabular}{r c l}
\Rev{$M_U$} & \Rev{$:=$} & \Rev{$\{(x, y) \mid (0 \leq x < n) \wedge (0 \leq y < n) \wedge (x = y)$\}}
\end{tabular}

\noindent
\Rev{Any set representation can be shown in the form of tensor computation over a boolean domain by replacing $\wedge$ with $*$ and $\vee$ with $+$. Therefore, the unique set can be represented as follows:}

\begin{tabular}{r c l}
\Rev{$M_U(x, y)$} & \Rev{$:=$} & \Rev{$(0 \leq x < n) * (0 \leq y < n) * (x = y)$}
\end{tabular}

\Rev{\smartpara{Redundancy map}} 
It provides the mapping from redundant elements to their corresponding unique elements and corresponds to capturing the redundancy pattern. \Rev{For example, the unique set and redundancy map for an $n \times n$ symmetric matrix $V$ are as follows:}

\begin{tabular}{r c l}
\Rev{$V_U(x, y)$} & \Rev{$:=$} & \Rev{$(0 \leq x < n) * (0 \leq y < n) * (x \leq y)$} \\
\Rev{$V_R(x, y, x', y')$} & \Rev{$:=$} & \Rev{$(0 \leq x < n) * (0 \leq y < n) * (x > y) * (x' = y) * (y' = x)$}
\end{tabular}

\noindent
\Rev{Here we are assuming that the lower triangular part of the matrix is the unique part, and the upper triangular part of it is repetitive. Variables $x, y$ represent the iteration space, and the mapping from the upper triangular to the lower triangular part is represented through variables $x', y'$. So this representation means $x$ and $y$ values must be swapped to have access to the unique elements.}

\Rev{\smartpara{Structure inference}}
\Rev{Now imagine the example of} sparse matrix-matrix element-wise multiplication \Rev{where the operands are the aforementioned matrices, $M$ and $V$. To perform this multiplication while leveraging the structure,} \stur{} applies the program reasoning rules~\cite{structtensor} to infer this computation's output structure and compressed iteration space. For the mentioned example, the \Rev{computation and the inferred} output structure \Rev{after simplification} are:

\begin{tabular}{r c l}
\Rev{$T(x, y)$} & \Rev{$:=$} & \Rev{$M(x, y) * V(x, y)$} \\
\Rev{$T_U(x, y)$} & \Rev{$:=$} & \Rev{$(0 \leq x < n) * (0 \leq y < n) * (x = y)$} \\
\Rev{$T_R(x, y, x', y')$} & \Rev{$:=$} & \Rev{$\emptyset$}
\end{tabular}

\Rev{\smartpara{Compressed tensor}
It is used for code generation and contains the structural information (the unique set and redundancy map) and the arithmetic operation over tensors using \stur{}'s syntax. The computation over sets and arithmetic is written using the same intermediate representation (\stur{}) for structure inference purposes but is treated separately in the code generation process. \structtensor{} internally separates the structural information and arithmetic parts. The structural information is used to generate loops while the arithmetic part generates the computational code inside the loop nests. For example, matrix $T$ will have the following compressed tensor:}

\begin{tabular}{r c l}
\Rev{$T_C(x, y)$} & \Rev{$:=$} & \Rev{$(0 \leq x < n) * (0 \leq y < n) * (x = y) * M(x, y) * V(x, y)$}
\end{tabular}

\noindent
\Rev{This means that in this computation, the iterators $x, y$ should iterate between $0$ and $n$ and must be equal to each other. The arithmetic operation that they perform inside the loop nests is $M(x, y) * V(x,y)$. Note that compressed tensor here refers to a tensor that has the structural information, hence, the iteration space can be compressed using this information. The tensor data layout itself is not compressed here. \structtensor{} generates the following code:

}

\vspace{-0.2cm}

\begin{lstlisting}[language=C++,keywordstyle=\color{blue},stringstyle=\color{red},commentstyle=\color{green}]
for (int x = 0; x < n; ++x) {
  int y = x;
  T(x, y) = M(x, y) * V(x, y);
}
\end{lstlisting}

\vspace{-0.2cm}

\noindent
\Rev{The first 2 lines are generated by the structural information and the third line is generated because of the arithmetic part of the compressed tensor.}

\smartpara{State-of-the-Art} \structtensor{}~\cite{structtensor} introduces the \stur{} intermediate language with the program reasoning rules to infer structures. As a final step, \structtensor{} provides C++ code that performs the computation over the compressed iteration space. \structtensor{} outperforms sparse and dense tensor algebra competitors by leveraging and propagating the compile-time-known structures throughout the computation. 

\smartpara{Limitations} Even though \structtensor{} infers the structure, it relies on an uncompressed dense representation of tensors that contains zero and repetitive elements. Alternatively, \structtensor{} requires the user to specify a compact data layout manually, but it is limited to only data layout for inputs. The outputs are still stored in an uncompressed dense format. However, \structtensor{} cannot capture this compressed output data layout. Furthermore, the generated C++ code relies on the low-level compiler (e.g., Clang or GCC) for further optimizations. Thus, one cannot benefit from the more advanced optimizations not supported directly by the compiler (e.g., parallelization). 

\systemname{} solves both limitations by relying on the mathematical foundation of the polyhedral model. First, it uses a novel symbolic indexing algorithm to densely pack the input and output tensors. Second, benefitting from the progressive lowering between intermediate languages (dialects) provided by the MLIR framework, it applies additional compiler optimizations, including parallelization and vectorization.





\begin{figure}
    \centering
    \includegraphics[width=\linewidth]{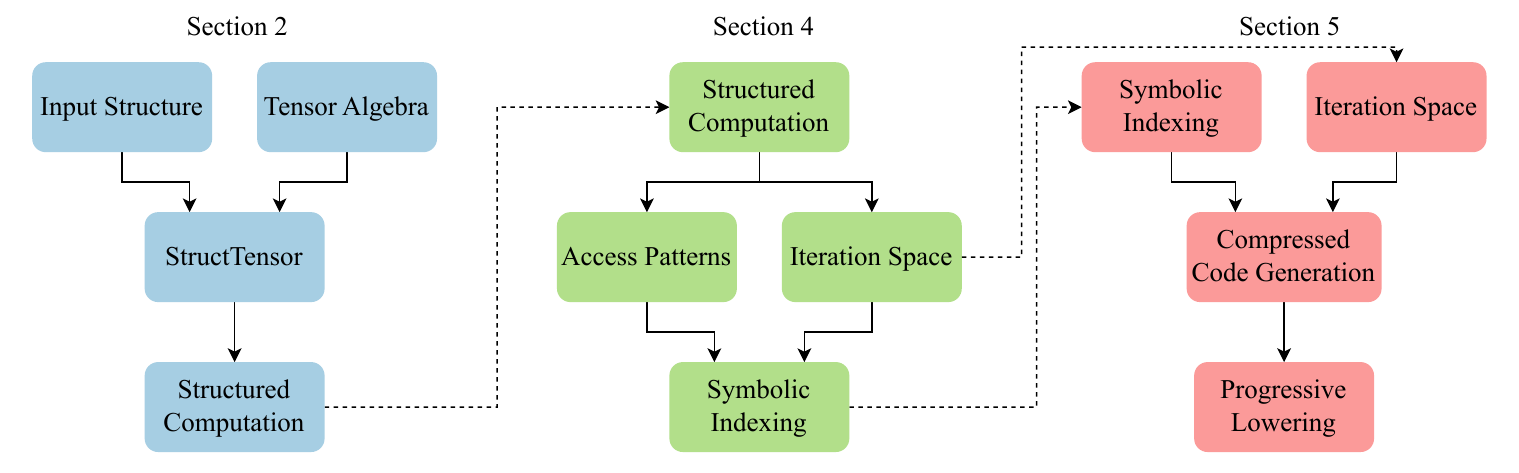}
    \caption{\systemname{} architecture overview.}
    \label{fig:architecture}
    \vspace{-0.5cm}
\end{figure}

\section{Overview}
\label{sec:overview}

In this section, we describe the overall architecture of \systemname{} (cf. Figure~\ref{fig:architecture}).


\begin{figure}[t]
\nonumber
\begin{equation}
\begin{split}
A(i, j, k) &:= B(i, j, l) * C(k, l)\\
C_U(k, l) &:= (0 \le k) * (k < P) * (0 \le l) * (l < Q)\\
B_U(i, j, l) &:= (i \le j) * (0 \le i) * (i < M) * (0 \le j) * (j < N) * (0 \le l) * (l < Q)\\
A_C(i, j, k) &:= C(k, l) * B(i, j, l) * (0 \le l) * (Q > l) \\
& * (i \le j) * (N > j) * (0 \le i) * (M > i) * (0 \le k) * (P > k)
\end{split}
\end{equation}
\vspace{-0.4cm}
\caption{\Rev{The TTM running example where the first tensor has an upper half cube structure. $A_C$ represents its optimized structured computation.}}
\label{listing:ttm_dp}
\vspace{-0.5cm}
\end{figure}

\smartpara{Input}
\systemname{} uses \stur{} as its intermediate language, as it
relies on \structtensor{}~\cite{structtensor} for the structure inference as the first step. \stur{} can express generalized Einsum expressions as a sum of products (SoP). As an example, Figure~\ref{listing:ttm_dp} shows the \stur{} representation of the tensor times matrix (TTM) computation with an upper half cube structure.

\smartpara{Structure inference} 
\systemname{} proceeds to propagate the input structure throughout the computation by applying program reasoning rules provided by \structtensor{}. The output structure and compressed computation are inferred during this stage. \systemname{} uses the inferred compressed domain by \structtensor{} as input. For the running example, the output structure is as follows:

\begin{tabular}{r c l}
     $A_U(i, j, k)$ & $:=$ & $B_U(i, j, l) * C_U(k, l)$
\end{tabular}

\smartpara{Structure Optimizations} 
\systemname{} improves the structured tensor computation by applying structure optimizations. \structtensor{} provides a set of optimizations such as inlining and logical simplifications (e.g., set idempotence and addition/multiplication identity optimizations) on the output structure and compressed computation.
A polyhedral tool such as isl~\cite{isl} can easily support and apply the aforementioned optimizations alongside further polyhedral set optimizations. Therefore, relying on polyhedral tools can reduce the development cost while increasing the robustness. For these reasons, \systemname{} relies on isl for further structure optimizations. Figure~\ref{listing:ttm_dp} shows the structured computation for the running example after applying structure optimizations.


\smartpara{Data Layout Compression} 
\label{sec:overview:compression}
\systemname{} leverages the statically known sparsity structure to assemble a densely packed buffer, aiming for cache-friendliness and vectorization opportunities. Our compression technique consists of a symbolic indexing function, mapping each original tensor index to a linear index of unique and non-zero values.
For example, the inferred compressed index of \Rev{$B$} in our running example Figure~\ref{listing:ttm_dp} is:
\Rev{
$$P_{B}(i,j,l) = \left(\left(\left(-\frac{1}{2} + N\right) Q \cdot i -\frac{1}{2} \cdot Q \cdot i^2\right) + Q \cdot j\right) + l$$
}

In this expression, the division operator is rational, $Q$ and $N$ correspond to the dimensions in Figure~\ref{listing:ttm_dp}.
Compression and decompression can then be achieved by iterating over the accessed indices and copying the data in and out of the compressed buffer.

\smartpara{Progressive Code Generation}
\systemname{} then proceeds to code generation. It leverages isl's Abstract Syntax Tree (AST) generation feature, taking a union of iteration domains and optional validity dependencies and returns an AST enforcing those conditions.
This AST is then used to generate code in the Affine dialect, a polyhedral-based intermediate representation defined in the MLIR framework. We drive the compilation from this IR down to executable code through MLIR's provided passes to generate low-level code implementing the used affine operations, simplify the polynomials computation and hoist each part as high as possible in the loop nest, and parallelize the computation by mapping to OpenMP runtime calls. Finally, the resulting LLVM IR is passed to Clang, which provides auto-vectorization and profits from the new opportunities to do so offered by this densely packed data layout representation.

\begin{figure}[t]
    \centering
    \begin{minipage}{\textwidth}
    \begin{multicols}{2}
        \begin{mlirln}
scf.parallel (%arg7) = (%c0) to (%7)
             step (%c1) {
  %8 = muli %arg7, %c-1 : index
  %9 = muli %8, %arg6 : index
  %10 = muli %arg7, %c2 : index
  %11 = muli %10, %arg4 : index
  %12 = muli %11, %arg6 : index
  %13 = addi %9, %12 : index
  %14 = muli %arg7, %arg7 : index
  %15 = muli %14, %c-1 : index
  %16 = muli %15, %arg6 : index
  %17 = addi %13, %16 : index
  %18 = muli %8, %arg3 : index
  %19 = muli %10, %arg3 : index
  %20 = muli %19, %arg4 : index
  %21 = addi %18, %20 : index
  %22 = muli %15, %arg3 : index
  %23 = addi %21, %22 : index
  scf.for %arg8 = %arg7 to %arg4 step %c1 {
    %24 = muli %arg8, %c2 : index
    %25 = muli %24, %arg6 : index
%26 = addi %17, %25 : index
%27 = muli %24, %arg3 : index
%28 = addi %23, %27 : index
scf.for %arg9 = %c0 to %arg6 step %c1 {
  %29 = muli %arg9, %c2 : index
  %30 = addi %26, %29 : index
  %31 = divui %30, %c2 : index
  %32 = muli %arg9, %arg3 : index
  scf.for %arg10 = %c0 to %arg3 step %c1 {
    %33 = load %alloc[%31] : memref<?xf64>
    %34 = addi %32, %arg10 : index
    %35 = load %alloc_0[%34] : memref<?xf64>
    %36 = muli %arg10, %c2 : index
    %37 = addi %28, %36 : index
    %38 = divui %37, %c2 : index
    %39 = load %alloc_1[%38] : memref<?xf64>
    %40 = mulf %35, %39 fastmath<fast> : f64
    %41 = addf %33, %40 fastmath<fast> : f64
    store %41, %alloc[%31] : memref<?xf64>
  }}}}
        \end{mlirln}
        \end{multicols}
    \end{minipage}
    \caption{The compressed kernel code only consists of cheap arithmetic operations, loops, and loads and stores on data. In particular, no index and offset values are loaded from memory such that all memory bandwidth is available for the actual compute workload}
    \label{listing:ttm_dp_mlir}
\end{figure}

\section{Data Layout Compression}
\label{sec:datalayout}
In this section, we present our method for data layout compression using a novel symbolic indexing algorithm, enabling a densely packed representation of the structured sparse tensor elements. This method relies on a symbolic and linear indexing of the values used in a computation. As opposed to well-established sparse data layouts, such as CSR, CSC, or COO, that incur access indirections and index storage overhead, our method uses direct symbolic index computations.
This direct indexing simultaneously reduces the global memory footprint and opens vectorization opportunities, two excellent properties when memory bandwidth is a significant constraint on the program efficiency, as is the case on many widespread contemporary hardware, such as CPUs and GPUs.

\systemname{} starts \Rev{the compression} process from a tensor computation expressed in \stur{} (e.g., Figure~\ref{listing:ttm_dp}). \systemname{} restricts itself to the \stur{} rules expressed with quasi-affine expressions \Rev{(affine expressions augmented with modulos)}. This enables the implementation to leverage mature tools to infer this compressed indexing efficiently.

\Rev{
To sidestep combinatorial complexity, \systemname{} only works with individual summands of a rule. Take for example the following rule:
$$O(j) := A(i,j) * (i = 1) * (0 \leq j < N) + A(i,j) * (i = 3) * (0 \leq j < N)$$
\systemname{} will first infer linear indices for $O(j)$ and $A(i,j)$ over the first region, where $i = 1$.
Only then does it move on to the $i = 3$ region.
In the second region, with $i = 3$, a disjoint region of $A$ is read from, so a second compressed buffer is inferred for this region. 
On the other hand, the same region of $O$ is being written to; \systemname{} thus "deduplicates" the compressed buffer by simply reusing the one infered for the first region. 

This design decision avoids compile-time costly generalization of the process and is found to work effectively on many motivated cases. However, in all generality, this yields two limitations.
First, some values might be duplicated in different compressed buffers, potentially leading to a suboptimal memory footprint reduction in some specific cases. Second, the output compression is only used when output regions are equal (as in the above example) or disjoint, as any other case would potentially lead to partial results being spread across multiple buffers, requiring further computation overhead to have the right values in each buffer.
}

\begin{algorithm}[t!]
    \begin{algorithmic}[1]
        \Function{SymbolicIndexing}{$\mathcal{D}_{\mathbf{p}}$, $T$}
            \State $\mathcal{A}_{\mathbf{p}} \gets$ \Call{AccessMapping}{$\mathcal{D}_{\mathbf{p}}, T$}
            \State $\mathcal{B}_{\mathbf{p},\mathbf{i}} \gets$ \Call{LocalOrdering}{$\mathcal{A}_{\mathbf{p}}, T$}
            \State $T_{index}(\mathbf{i}, \mathbf{p}) \gets$ \Call{SymbolicCount}{$\mathcal{B}_{\mathbf{p},\mathbf{i}}$}
            \State \Return \Call{PiecewiseFusion}{$T_{index}$}
        \EndFunction
    \end{algorithmic}
\caption{\Rev{Symbolic Indexing Algorithm}}
\label{algo:symbolic_indexing}

\end{algorithm}

\smartpara{Symbolic Indexing Algorithm}
At this stage, \systemname{} uses our symbolic indexing algorithm (Algorithm~\ref{algo:symbolic_indexing}) to infer a compressed indexing function.
The algorithm takes two inputs. The first input is $\mathcal{D}_\mathbf{p}$ which is an ordered iteration space, where the parameter $\mathbf{p}$ defines symbols such as dimension information. In other words, $\mathcal{D}_\mathbf{p}$ represents the set of iterator values with which the tensor access $T$ is used. By ordered we mean a sequential execution of the computation. The second input denoted by $T$, is the considered tensor access in the form of a function from an iteration to the indices used in this access.
The described process in Algorithm~\ref{algo:symbolic_indexing} is illustrated using buffer A of our example in Figure~\ref{listing:ttm_dp}.

\Rev{
\label{sec:compile:compression}
}

\smartpara{Iteration space extraction}
To apply Algorithm~\ref{algo:symbolic_indexing}, we define the iteration space of the summand at hand, taking all comparisons of the considered \stur{} summand.
For our running example in Figure~\ref{listing:ttm_dp}, the iteration space is:

\vspace{-0.2cm}


$$\mathcal{D}_{Q,N,M,P} := \{ (i,j,k,l) \in \mathbb{Z}^4 \mid 0 \leq i < M, i \leq j < N, 0 \leq k < P, 0 \leq l < Q \}$$

\noindent
Here, symbols correspond to the boundary information of dimensions ($Q,N,M,P$). By restricting \stur{} to affine expressions, one can observe that this set is a polyhedron by construction.

\Rev{
Because we define the iteration space from a high-level representation, we control the order of the dimensions; we simply order them based on their appearance in the rule, ensuring the first iterators are used to index the output, with the last one of them indexing contiguous elements.
}






\smartpara{Access mapping} For each buffer, we define its access map using the access variables used in the \stur{} expression. By construction, those are affine maps. The corresponding map for buffer $B$ in Figure~\ref{listing:ttm_dp} \Rev{is as follows}:
$$a_B(i,j,k,l) = (i,j,l)$$
$$\mathcal{A}_{B,\overrightarrow{p}} := a_B(\mathcal{D}_{\overrightarrow{p}}) = \{ \overrightarrow{j} \in \mathbb{Z}^m \mid \exists i \in \mathbb{Z}^n : a_B \cdot \overrightarrow{i} = \overrightarrow{j}, \mathcal{C} \cdot \overrightarrow{i} + \overrightarrow{c} \geq 0 \}$$

\Rev{
\noindent
We apply this mapping on the above iteration space, yielding the set of accessed indices:
\begin{equation}
    \begin{split}
\mathcal{A}^B_{Q,N,M,P} = a_B(\mathbf{\mathcal{D}_{Q,N,M,P}}) &= \{(i,j,l) \in \mathbb{Z}^3 \mid 0 \leq i < M, i \leq j < N, 0 \leq l < Q\}
    \end{split}
\end{equation}

\vspace{-0.2cm}
}

    
    
\Rev{
\smartpara{Preceding Accesses}
We proceed to define an order on this accessed domain, using the access map to respect the iteration order. This order is similar to the lexicographic order, only using indices in their iteration domain order. For the map $a(i,j,k,l) = (k,j,l)$, this order will be: $$a(i',j',k',l') < a(i,j,k,l) \iff j' < j \vee j' = j \wedge k' < k \vee j' = j \wedge k' = k \wedge l' < l$$

We use the set of iterators used in the access map in their initial order:


Applying this order to the accessed domain, we define the preceding access domain as the set of all accessed indices preceding a symbolic index. For our running examples, this yields:
}
\begin{equation*}
    \begin{split}
\mathcal{P}^B_{Q,N,M,P,i,j,l} &= \{(i',j',l') \in \mathcal{A}^B_{Q,N,M,P} \mid i' < i\}\\ &\cup \{(i',j',l') \in \mathcal{A}^B_{Q,N,M,P} \mid i' = i \wedge j' < j\}\\ 
&\cup \{(i',j',l') \in \mathcal{A}^B_{Q,N,M,P} \mid i' = i \wedge j' = j \wedge l' < l \}
    \end{split}
\end{equation*}



\smartpara{Access domain linearization} Finally, we can synthesize a mapping from a current iteration to the desired compressed linear index by counting the number of effective accesses preceding this iteration, that is, counting the number of elements of $\mathcal{A}_{\mathbf{p},\mathbf{i}}$. 
The Barvinok algorithm~\cite{barvinok} is a method for counting the number of integer points within a parametric polyhedron, returning a piecewise quasi-polynomial \Rev{(a generalization of polynomials having periodic coefficients) of the polyhedron's parameters.

We can thus use this algorithm, implemented in isl, on the preceding access domain and obtain a linear indexing of the accessed domains element, following our chosen order:}
$$P_B = \#\mathcal{P}^B_{\mathbf{p},\mathbf{i}}$$

\begin{figure}
    \centering
    \resizebox{0.48\linewidth}{!}{
    \begingroup
        \fontsize{16pt}{8pt}\selectfont
        \input{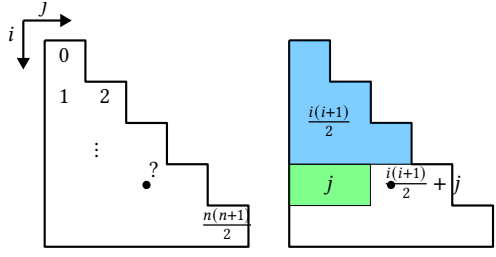}
    \endgroup
    }
    \caption{Illustration of the affine symbolic indexing method, ensuring a dense packing and \Rev{memory compression}.}
    \label{fig:indexing_polynomial}
\end{figure}

\begin{algorithm}[t!]
    \begin{algorithmic}[1]
        \Function{PiecewiseFusion}{$T_{index}$}
        \ForAll{$(D,P)$ in $T_{index}$}
             \ForAll{$(D',P')$ in $T_{index} - (D,P)$}
                \If{$(P-P')\vert_{D'} = 0$}
                    \State $D \gets D \cup D'$
                    \State $T_{index} \gets T_{index} - (D',P')$
                \EndIf
            \EndFor
        \EndFor
        \State \Return $T_{index}$
        \EndFunction
        
    \end{algorithmic}
\caption{\Rev{Piecewise QuasiPolynomial Fusion Algorithm}}
\label{algo:piecewise_fusion}
\end{algorithm}

\Rev{

\smartpara{Piecewise Polynomial Simplification} In general, the counting algorithm yields a \emph{piece-wise} quasi-polynomial, which could have a different expression on distinct subsets of the counted set. In practice, it often yields special cases at boundaries, that are not desirable for simple code generation. For example, when counting the preceding accesses for a triangle as illustrated in Figure~\ref{fig:indexing_polynomial}, the resulting piecewise polynomial is:
\begin{equation}
P =
\begin{cases}
  \frac{1}{2}i + \frac{1}{2}i^2 + j & \text{if } j > 0\\
  \frac{1}{2}i + \frac{1}{2}i^2 & \text{if } j = 0
\end{cases}
\end{equation}

We apply a piecewise polynomial fusion algorithm presented in Algorithm~\ref{algo:piecewise_fusion} to simplify the expression of this polynomial. Let us apply this algorithm to our example. We name the two regions as $D$ and $D'$, and their corresponding polynomials as $P$ and $P'$: $P = \frac{1}{2}i + \frac{1}{2}i^2 + j$, $P' = \frac{1}{2}i + \frac{1}{2}i^2$. Following line 4 of Algorithm~\ref{algo:piecewise_fusion}, we have $P-P' = j$. Restricting this difference to $D' : j = 0$, we have $(P-P')\vert_{D'} = 0$; in other words, $P$ is equal to $P'$ on the latter's domain. We thus remove the special case on this domain (line 6), only keeping $P$ on the union of $D$ and $D'$, yielding a single polynomial expressive enough for the whole buffer:
$$P = \frac{1}{2}i + \frac{1}{2}i^2 + j$$
}

The computed compressed index for buffer $A$ in Figure~\ref{listing:ttm_dp} is as follows:
$$P_{B}(i,j,l) = \left(\left(\left(-\frac{1}{2}Q + Q \cdot N\right) \cdot i -\frac{1}{2} Q \cdot i^2\right) + Q \cdot j\right) + k$$

In this example, one is expressive enough for the whole buffer; we can thus safely ignore the conditions, knowing by construction we will not index any value out of those bounds.

An interesting side-effect of our design is that the generated piece-wise quasi-polynomials are readily expressed in a form where the slowest varying indices are evaluated the deepest in the expression, making it amenable to effective loop-invariant code hoisting. For example, in $P_B$, the most product-intensive term ($-\frac{1}{2}Q + Q \cdot N$) can be hoisted of all loops in the computation, as it only depends on the sizes of the tensors.

\begin{figure}[t]
\begin{minipage}{0.45\linewidth}
\begin{mlir}
func.func @plus10(%x : index) -> index {
  %10 = arith.constant 10 : index
  %ret = arith.addi %x, %10 : index
  func.return %ret
}
\end{mlir}
\end{minipage}
\vspace{-0.3cm}
\caption{Example function in MLIR IR.}
\label{listing:mlir_example}
\vspace{-0.5cm}
\end{figure}

\Rev{
As our experimental results show, our symbolic indexing algorithm significantly reduces the memory footprint. Furthermore, we observed performance improvements thanks to improved cache locality and enabling vectorization opportunities. Finally, we have not observed significant performance overhead caused by the index computation time on the selected motivated cases. 
}

\section{Progressive Code Generation}
\label{sec:compile:codegen}

Finally, \stur{} starts generating code using the MLIR framework. It has been chosen as a compiler for this work for its ability to express high-level Intermediate Representation (IR) and lower it progressively to LLVM IR, a low-level IR that the CLang compiler can compile down to binary. This enables MLIR to provide many optimizations at a high level, where the information needed to apply them safely is still present. 

\subsection{MLIR Background}
MLIR \cite{mlir} is a compiler framework successfully applied in the development of compilers in deep learning~\cite{pytorch, tensorflow}, fully-homomorphic encryption~\cite{SyFER2020}, and hardware design~\cite{eldridge2021mlir, majumder2021hir}.
It also opens opportunities, such as compiling
our high-level IR down to GPU kernels or other targets rather than generating specialized code and reinventing the wheel, but this has been left as future work.

MLIR IR elementary objects are Operations, Values, Regions, and Attributes. Values represent runtime values and are constrained by the Static Single-Assignment (SSA) form: they can only be assigned once, at definition time.
Operations may take values as input and may define new values as output. They can be augmented through attributes, representing various compile-time information.
Finally, Operations may contain Regions, which include lists of Operations and possibly entry arguments: Values defined in the region and taking their runtime value from the operation.

An example is given in Figure~\ref{listing:mlir_example}. \mlirinline{func.func} is an Operation representing a function definition. It has a Region representing its body and a function type.
In its region, \mlirinline{arith.constant} is an operation only defining a value from a compile-time known value and type. \mlirinline{arith.addi} takes this value and the function's argument and returns its sum in a new value. Finally, \mlirinline{func.return} only consumes this value to represent returning it from the function.

MLIR implements this idea of mixing levels of abstraction through dialects: collections of operations and types that represent some level of abstraction but can be combined in a program to allow progressive lowering of different parts in a controllable manner. Operations names are prefixed by their dialect's name: in Figure~\ref{listing:mlir_example}, arith and func were used.

\definecolor{color1}{HTML}{a6cee3}
\definecolor{color2}{HTML}{1f78b4}
\definecolor{color3}{HTML}{b2df8a}
\definecolor{color4}{HTML}{33a02c}
\definecolor{color5}{HTML}{fb9a99}
\definecolor{color6}{HTML}{e31a1c}
\definecolor{colorm1}{HTML}{cacaca}
\definecolor{colorm2}{HTML}{827b7b}

\newcommand{\codemapa}[1]{\smash{\setlength{\fboxsep}{1.5pt}{\colorbox{color1}{#1}}}}
\newcommand{\codemapb}[1]{\smash{\setlength{\fboxsep}{1.5pt}{\colorbox{color2!80}{#1}}}}
\newcommand{\codemapc}[1]{\smash{\setlength{\fboxsep}{1.5pt}{\colorbox{color3}{#1}}}}
\newcommand{\codemapd}[1]{\smash{\setlength{\fboxsep}{1.5pt}{\colorbox{color4}{#1}}}}
\newcommand{\codemape}[1]{\smash{\setlength{\fboxsep}{1.5pt}{\colorbox{color5}{#1}}}}
\newcommand{\codemapf}[1]{\smash{\setlength{\fboxsep}{1.5pt}{\colorbox{color6!80}{#1}}}}
\newcommand{\codemapg}[1]{\smash{\setlength{\fboxsep}{1.5pt}{\colorbox{colorm1}{#1}}}}
\newcommand{\codemaph}[1]{\smash{\setlength{\fboxsep}{1.5pt}{\colorbox{colorm2}{#1}}}}

\begin{figure}[t]
\begin{minipage}{0.49\columnwidth}
\begin{mlir}
mark: parallel
iter: |\codemapa{i}|, init: |\codemape{0}|, cond: |\codemapc{i < min(N,M)}|, inc: 1
  iter: |\codemapa{j}|, init: |\codemape{i}|, cond: |\codemapc{j < N}|, inc: 1
    iter: |\codemapa{k}|, init: |\codemape{0}|, cond: |\codemapc{k < P}|, inc: 1
      iter: |\codemapa{l}|, init: |\codemape{0}|, cond: |\codemapc{l < Q}|, inc: 1
        S0[i, j, k, l]
.
\end{mlir}
\end{minipage}
\begin{minipage}{0.09\columnwidth}
$\longrightarrow$
\end{minipage}
\begin{minipage}{0.4\columnwidth}
\begin{mlir}
affine.parallel |\codemapa{%i}| = |\codemape{0}| to |\codemapc{min(%N, %M)}| {
  affine.for |\codemapa{%j}| = |\codemape{%i}| to |\codemapc{%N}| {
    affine.for |\codemapa{%k}| = |\codemape{0}| to |\codemapc{%P}| {
      affine.for |\codemapa{%l}| = |\codemape{0}| to |\codemapc{%Q}| {
        S0(i, j, k, l);
}}}}
\end{mlir}
\end{minipage}
\caption{The translation of ISL AST to MLIR's Affine loops.}
\label{listing:ast_example}
\end{figure}

\subsection{Affine code generation}

\Rev{
\smartpara{Parallelization} 
\systemname{} drives ISL to generate an AST expressing parallelism over the outermost index if it is not a reduction.
We believe that parallel code generation is a benefit of our densely packed data layout; while similar compaction could be achieved by simple linear indices increments judiciously placed in the loop nest, those would effectively result in loop-carried dependencies, and necessitate further analysis or treatment to parallelize correctly.
\systemname{}'s symbolic indexing, on the other hand, only depends on current iterators and symbols of the computation; once inferred, it stays a correct mapping however loops are parallelised or further transformed.
Thus, parallelism, loop transformations, and compression usage in the generated code become completely orthogonal.
Our experiments demonstrate the effectiveness of our compression scheme with simple parallelization (cf. Section~\ref{sec:experiment:e2e}).
}

\smartpara{Control Flow Code Generation}
The first step is obtaining an Abstract Syntax Tree (AST) scanning the iteration domain from ISL; an example is in Figure~\ref{listing:ast_example}. By construction, this AST is guaranteed to have affine bounds and increments. \systemname{} uses MLIR's affine dialect, which expresses affine loops, affine conditions, and affine accesses, to generate IR at the closest abstraction level from its internal one.
The main benefit of this approach is that \systemname{} does not need to re-implement the basics: it simply translates affine expressions from ISL's to MLIR's representation and defines loops using those. MLIR is equipped to lower those expressions to arithmetic computations and has advanced infrastructure to simplify those computations at a lower level. 
Finally, \systemname{} generates code corresponding to each statement. For the computation expressed in the \stur{} language, that means loading each designated value from the right-hand side, multiplying them, and adding that product to the loaded left-hand side value. \Rev{Memory accesses are translated in two steps to simplify the usage of the compressed indexing polynomial, as explained below.}

\smartpara{Uncompressed Compute Code Generation}
The first step is to apply ISL's \emph{schedule}, an affine mapping from the current iteration to the values to pass to the statement. In the example given in Figure~\ref{listing:ttm_dp_mlir}, this schedule is the identity $\theta(i,j,k,l)=(i,j,k,l)$.

\begin{figure}[t]
\begin{minipage}{0.2\columnwidth}
\begin{mlir}
S0[|\codemapa{i}|,
   |\codemapb{j}|,
   |\codemapc{k}|,
   |\codemapd{l}|]
\end{mlir}
\end{minipage}
\begin{minipage}{0.04\columnwidth}
$\longrightarrow$
\end{minipage}
\vspace{2.2pt}
\begin{minipage}{0.68\columnwidth}
\begin{mlir}
|\codemapa{\%si}| = affine.apply affine_map<|\codemapa{(i,j,k,l)->(i)}|>(\%i, \%j, \%k, \%l)
|\codemapb{\%sj}| = affine.apply affine_map<|\codemapb{(i,j,k,l)->(j)}|>(\%i, \%j, \%k, \%l)
|\codemapc{\%sk}| = affine.apply affine_map<|\codemapc{(i,j,k,l)->(k)}|>(\%i, \%j, \%k, \%l)
|\codemapd{\%sl}| = affine.apply affine_map<|\codemapd{(i,j,k,l)->(l)}|>(\%i, \%j, \%k, \%l)
\end{mlir}
\end{minipage}
\begin{minipage}{0.2\columnwidth}
\begin{equation*}
    \begin{split}
&\codemape{A(i, j, k)} :=\\&\codemapf{C(k, l)} * \codemapg{B(i, j, l)}
    \end{split}
\end{equation*}
\end{minipage}
\begin{minipage}{0.04\columnwidth}
$\longrightarrow$
\end{minipage}
\begin{minipage}{0.68\columnwidth}
\begin{mlir}
|\%|lastval = affine.load |\codemape{\%A[\%si, \%sj, \%sl]}| : memref<?x?x?xf64>
|\%|factor1 = affine.load |\codemapf{\%C[\%sk, \%sl]}| : memref<?x?xf64>
|\%|factor2 = affine.load |\codemapg{\%B[\%si, \%sj, \%sk]}| : memref<?x?x?xf64>
|\%|prod = arith.mulf |\%|factor1, |\%|factor2 : f64
|\%|newval = arith.addf |\%|lastval, |\%|prod : f64
affine.store |\codemape{\%A[\%si, \%sj, \%sl]}| : memref<?x?x?xf64>
\end{mlir}
\end{minipage}
\caption{ISL's schedule is naturally translated to affine maps application and the input computation to accesses and arithmetic.}
\label{listing:schedule_translation}
\end{figure}

MLIR's affine dialect provides an operation to apply a given affine mapping to a list of indices, \mlirinline{affine.apply}, so we use it to define new values corresponding to the scheduled access variables. It then remains to generate the product and accumulation code from the corresponding accesses in the input \stur{} rule, using MLIR's basic arithmetic operations. THis is all illustrated in Figure~\ref{listing:schedule_translation}


\begin{figure}[t]
\begin{minipage}{0.2\columnwidth}
\begin{equation*}
    \begin{split}
&\codemape{A(i, j, k)} :=\\&\codemapf{C(k, l)} * \codemapg{B(i, j, l)}
    \end{split}
\end{equation*}
\end{minipage}
\begin{minipage}{0.04\columnwidth}
$\longrightarrow$
\end{minipage}
\begin{minipage}{0.68\columnwidth}
\begin{mlir}
|\codemape{%A\_C\_i}| = func.call |\codemape{@A\_P(%si, %sj, %sl, %Q, %N)}|
|\codemapg{%B\_C\_i}| = func.call |\codemapg{@B\_P(%si, %sj, %sk, %Q, %N)}|
|\codemapf{%C\_C\_i}| = func.call |\codemapf{@C\_P(%sk, %sl, %N)}|
%lastval = memref.load |\codemape{%A\_C[%A\_C\_i]}| : memref<?x?x?xf64>
%factor1 = memref.load |\codemapf{%C\_C[%C\_C\_i]}| : memref<?x?xf64>
%factor2 = memref.load |\codemapg{%B\_C[%B\_C\_i]}| : memref<?x?x?xf64>
%prod = arith.mulf %factor1, %factor2 : f64
%newval = arith.addf %lastval, %prod : f64
memref.store |\codemape{%A\_C[%A\_C\_i]}| : memref<?x?x?xf64>
\end{mlir}
\end{minipage}
\caption{Using our symbolic indexing, compressed indexing is natural to switch to from dense indexing.}
\label{listing:compressed_translation}
\end{figure}

\smartpara{Compressed Compute Code Generation}
\Rev{
Because the inferred compression polynomial is indexing a linear contiguous buffer by directly mapping indices of the original tensor to the corresponding compressed index, generating compressed compute code can be done independently of any control flow; it suffices to generate accesses to the compressed buffers using the inferred polynomial. This is illustrated in Figure ~\ref{listing:compressed_translation}, where those polynomials are represented as function calls.
}
The attentive reader might notice that we switched from \mlirinline{affine.load} to \mlirinline{memref.load}. Both representing a load operation, the affine variant restricts the access indices to affine expressions of iterators and symbols. Our usage of quasipolynomials breaks this constraint.
Here, MLIR's capacity to mix abstractions is helping because this doesn't forbid the usage of affine loops and maps where still applicable.


\subsection{Progressive Lowering}

\Rev{
The final step is progressively translating and optimizing this high-level representation to LLVM IR through MLIR. The result of the optimizations is illustrated in Figure~\ref{listing:ttm_dp_mlir}. The critical steps are as follows (cf. Section~\ref{sec:experiment} for the exact pipeline).
}

\smartpara{Function inling} The first optimization \systemname{} applies is MLIR's function inlining, \mlirinline{inline}. This is done to expose the quasi-polynomials arithmetic directly in the loop nest, allowing to share computations and hoisting of different parts as high as possible, amortizing their theoretical computational cost.

\smartpara{Lowering Affine} The first conversion to execute is from affine operations to arithmetic operations, implementing them through the \mlirinline{lower-affine} pass exposed by MLIR. This will expand all affine map applications, loop bounds, conditions, and accesses to as many basic arithmetic operations as necessary to correctly implement them.

\smartpara{Generic Optimizations} CSE is directly applied to the resulting IR. Amongst the different affine and quasipolynomial expressions used in the computation, all expressed on the same iterators and symbols, there are potentially a lot of common sub-expressions. By applying it directly at this level, we allow more opportunities than expecting another lower-level compiler to analyze simplification safety again.
The subsequent critical optimization is \mlirinline{canonicalization}, a collection of individual standard optimizations. In this case, we are primarily interested in its constant folding in case any opportunity is open.

\smartpara{Loop Invariant Code Motion} Finally, \systemname{} uses \mlirinline{loop-invariant-code-motion} to hoist all arithmetic operations as high as possible in the loop nest, leaving only minimal iteration arithmetic in the innermost loop to maximize auto-vectorization opportunities.

\section{Experimental Results}
\label{sec:experiment}

In this section, we experimentally evaluate our system on a set of application-driven selected kernels. We study the following questions:
\begin{itemize}[leftmargin=*]
	\item How does \systemname{} perform compared to the state-of-the-art tensor algebra frameworks in single- and multi-threaded environments?
	\item How does the automatic memory compression impact the allocated memory footprint? How do different compression levels improve performance?
        \item Can \systemname{} support complicated structures (e.g., block butterfly factor matrix~\cite{ChenDLY0RR22})? How does the structure's complexity impact the compression, run time, and compilation time?
        \item How does \systemname{} scale on various numbers of threads in comparison to state-of-the-art tensor algebra frameworks?
	\item What is the impact of applying various optimizations compared to the \structtensor{}? Can state-of-the-art backends (e.g., Polygeist) recover \systemname{}'s optimizations{}?
\end{itemize}

\subsection{Experimental Setup}

\begin{table}[t]
\caption{\Rev{Information on the kernels used for the end-to-end benchmarks.}}
\label{tbl:kernels_info}
\vspace{-0.2cm}
\scriptsize
\begin{tabular}{|c|c|c|c|}
\hline
\textbf{Kernel} & \textbf{Structure of $B$} & \Rev{\textbf{Name}} & \Rev{\textbf{Compile Time (s)}} \\
\hline
\textbf{TTM} & Diagonal (plane) & \Rev{TTM\_DP}  & \Rev{0.31} \\
 & Fixed $j$ & \Rev{TTM\_J}  & \Rev{0.31} \\
$A(i, j, k) := B(i, j, l) * C(k, l)$ & Upper half cube (UHC) & \Rev{TTM\_UT}  & \Rev{0.35} \\
\hline
\textbf{THP}  & Diagonal (plane) & \Rev{THP\_DP}  & \Rev{0.32}\\
 & Fixed $i$ & \Rev{THP\_I}  & \Rev{0.30} \\
$A(i, j, k) :=$ $B(i, j, k) * C(i, j, k)$ & Fixed $j$ & \Rev{THP\_J}  & \Rev{0.30} \\
\hline
\textbf{MTTKRP} & Diagonal & \Rev{MTT\_D}  & \Rev{0.30} \\
 & Fixed $j$ \& UHC & \Rev{MTT\_JUT}  & \Rev{0.33} \\
$A(i, j) := B(i, k, l)$ $ * $ $ C(k, j) * D(l, j)$ & Fixed $j$ & \Rev{MTT\_J}  & \Rev{0.33}\\
\hline
\textbf{SpMV} & Leslie & \Rev{SpMV\_L}  & \Rev{0.30} \\
 & Upper triangular & \Rev{SpMV\_UT}  & \Rev{0.30} \\
$A(i) := B(i, j) * C(j)$ & Diagonal & \Rev{SpMV\_D}  & \Rev{0.28} \\
\hline
\Rev{\textbf{Polynomial Regression Degree-D}} & \Rev{Dense, First Dimension Size = 100000} & \Rev{PR2}  & \Rev{0.47} \\
\Rev{$A^2, \ldots, A^{2D}$ where} & \Rev{Dense, First Dimension Size = 10000} & \Rev{PR3}  & \Rev{1.76}\\
\Rev{$A^{k}(i_1, \ldots, i_k) :=$ $B(x, i_1) * \ldots * B(x, i_k)$} & \Rev{Dense, First Dimension Size = 2000} & \Rev{PR4}  & \Rev{43.18}\\
\hline
\end{tabular}
\end{table}

We use a server running Ubuntu 22.04 LTS equipped with a 10-core 2.2 GHz Intel Xeon Silver 4210 CPU with one thread per core and 220 GB of main memory. The server has 32 KB, 1 MB, and 13.8 MB of level 1, 2, and 3 cache, respectively. All C++ code is compiled with Clang 18.0.0 using the following flags:
\begin{lstlisting}[basicstyle=\tiny]
 -std=c++17 -O3 -fopenmp -ffast-math -ftree-vectorize -march=native -mtune=native
\end{lstlisting}
MLIR code is lowered with the pass pipeline:

\begin{lstlisting}[basicstyle=\tiny]
builtin.module(canonicalize, cse, affine-expand-index-ops, lower-affine, finalize-memref-to-llvm, 
 canonicalize, cse, loop-invariant-code-motion, convert-arith-to-llvm, convert-scf-to-openmp, canonicalize, 
 convert-scf-to-cf,convert-func-to-llvm,convert-cf-to-llvm,convert-openmp-to-llvm,reconcile-unrealized-casts)
\end{lstlisting}
\systemname{} does not use tiling, as we did not observe a major benefit (cf. Section~\ref{sec:experiment:opt}).

As a competitor, we use Polygeist~\cite{polygeistPACT} to activate out-of-the-box MLIR code generation, polyhedral optimizations (e.g., tiling), and parallelization on top of 
\structtensor{}, since \structtensor{} lacks such optimizations as it is. 
Polygeist is shown to have a better single- and multi-threaded performance~\cite{polygeistPACT} compared to other tools such as Pluto~\cite{pluto} and Polly~\cite{polly}. 
Polygeist transforms a C code generated based on the inferred structure by \structtensor{} to MLIR with the following lowering pipeline:

\begin{lstlisting}[basicstyle=\tiny]
--canonicalize --cse --affine-expand-index-ops --affine-parallelize --affine-loop-tile --affine-loop-
invariant-code-motion --canonicalize --cse --lower-affine --loop-invariant-code-motion --convert-scf-to-
openmp --convert-polygeist-to-llvm --canonicalize --cse --loop-invariant-code-motion --canonicalize --cse
\end{lstlisting}

The lowered code goes through Clang with identical flags to generate executable binary files. For all experiments, the average run time of three runs is reported, as the numbers were stable. All experiments are evaluated using a warm cache. Cache counters are measured using PAPI 7.1~\cite{PAPI}.
As the competitors, we use the latest version of TACO\footnote{https://github.com/tensor-compiler/taco/tree/2b8ece4c230a5f}~\cite{kjolstad:2017:taco}, NumPy 1.26.0~\cite{numpy}, PyTorch 2.0~\cite{pytorch}, and TensorFlow 2.14.0~\cite{tensorflow} with the XLA backend. All the Python frameworks are run using Python 3.10.13. A dense implementation of the kernels does not leverage any structure and has been shown to perform worse than both sparse and dense tensor algebra frameworks~\cite{structtensor}; hence, we do not evaluate \systemname{} against it. 

Table~\ref{tbl:kernels_info} shows kernel information (including the ones used in \structtensor{}~\cite{structtensor}) with their structure, compilation time (time required for generating MLIR code using \systemname{}), and the name we used for them on the plots. This experimental setup has been used previously for \structtensor{}~\cite{structtensor}, a state-of-the-art structured tensor compiler. All the experiments except for the three kernels SpMV\_D, SpMV\_L, and MTT\_D fit in the cache.
Table~\ref{tbl:kernels} lists the kernels used in our experiments and their input structure representation. For TACO, we picked the best data layout according to the input structure for each kernel (cf. Table~\ref{tbl:kernels}). 

\begin{table}[t]
\caption{Tensor kernels for \systemname{} evaluation. In \Rev{MTT\_J}, $j$ is fixed for $D$. \Rev{In MTT\_D, $l = j$ too.}}
\label{tbl:kernels}
\vspace{-0.2cm}
\scriptsize
\begin{tabular}{|c|c|c|}
\hline
\textbf{Kernel} & \textbf{$B_U$ in \stur{}} & \Rev{\textbf{TACO Data Layout}} \\
\hline
\Rev{TTM\_DP} & $(0 \leq i < n_i) * (i = j) * (0 \leq l < n_l)$   & \Rev{$(D, S, D) * (D, D) \rightarrow (D, S, D)$}\\
\Rev{TTM\_J} & $(0 \leq i < n_i) * (j = J) * (0 \leq l < n_l)$ & \Rev{$(D, S, D) * (D, D) \rightarrow (D, S, D)$}\\
\Rev{TTM\_UT} & $(0 \leq i < n_i) * (i \leq j < n_j) * (0 \leq l < n_l)$ & \Rev{$(D, S, D) * (D, D) \rightarrow (D, S, D)$} \\
\hline
\Rev{THP\_DP} & $(0 \leq i < n_i) * (i = j) * (0 \leq l < n_l)$ & \Rev{$(D, S, D) * (D, D, D) \rightarrow (D, S, D)$} \\
\Rev{THP\_I} & $(i = I) * (0 \leq j < n_j) * (0 \leq l < n_l)$ & \Rev{$(S, D, D) * (D, D, D) \rightarrow (S, D, D)$} \\
\Rev{THP\_J} & $(0 \leq i < n_i) * (j = J) * (0 \leq l < n_l)$ & \Rev{$(D, S, D) * (D, D, D) \rightarrow (D, S, D)$}\\
\hline
\Rev{MTT\_D} & $(i = k = l) * (0 \leq i < n_i)$ & \Rev{$(D, S, S) * (D, D) * (D, S) \rightarrow (D, S)$} \\
\Rev{MTT\_JUT} & $(0 \leq i < k) * (0 \leq k < n_k) * (0 \leq l < n_l)$ & \Rev{$(D, S, D) * (D, D) * (D, S) \rightarrow (D, S)$} \\
\Rev{MTT\_J} & $(0 \leq i < n_i) * (0 \leq k < n_k) * (0 \leq l < n_l)$ & \Rev{$(D, D, D) * (D, D) * (D, S) \rightarrow (D, S)$}  \\
\hline
\Rev{SpMV\_L} & $(i = 0) * (0 \leq j < n_j) + (1 \leq i < n_i) * (j=i-1)$  & \Rev{$(D, S) * (D) \rightarrow (D)$}\\
\Rev{SpMV\_UT} & $(0 \leq i < n_i) * (i \leq j < n_j)$ & \Rev{$(D, S) * (D) \rightarrow (D)$} \\
\Rev{SpMV\_D} & $(0 \leq i < n_i) * (i = j)$ & \Rev{$(D, S) * (D) \rightarrow (D)$} \\
\hline
\end{tabular}
\end{table}

\subsection{End-to-End Experiments}
\label{sec:experiment:e2e}

In this subsection, we evaluated the performance of \systemname{} against the best alternative of \structtensor{} (w/ and w/o manual input data layout compression), TACO with the best structure for each kernel, and the best run time of Polygeist applied on the output of \structtensor{} (w/ and w/o manual input and output data compression, w/ and w/o tiling) on all the kernels (cf. Table~\ref{tbl:kernels}) in single- and multi-threaded environments. 



\begin{figure}
\setlength\tabcolsep{.5pt}
    \centering
    \begin{tabular}{c}
         \includegraphics[width=0.99\linewidth]{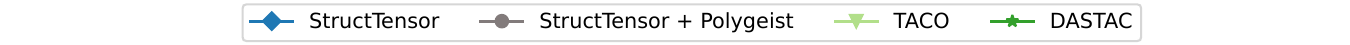}
    \end{tabular}
    \vspace{0.01\linewidth}
    \begin{tabular}{ccc}
         \includegraphics[width=0.25\linewidth,height=2.5cm]{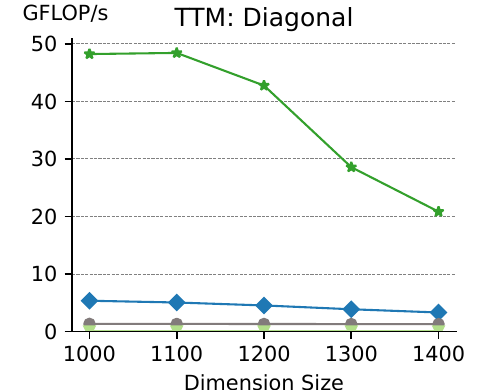} \hspace{0.045\linewidth} \vspace{0.01\linewidth} & \includegraphics[width=0.25\linewidth,height=2.5cm]{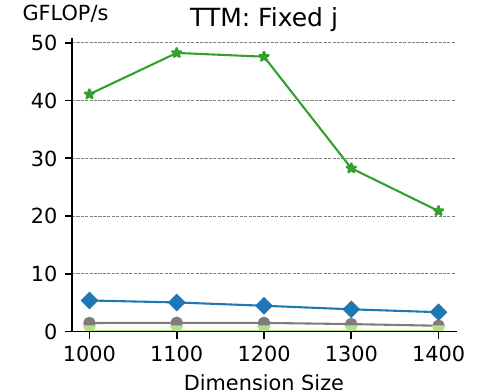} \hspace{0.045\linewidth} \vspace{0.01\linewidth} & \includegraphics[width=0.25\linewidth,height=2.5cm]{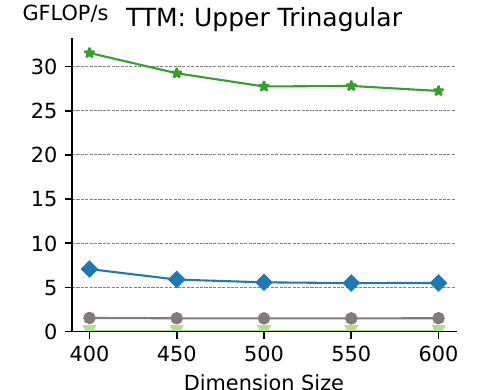}\vspace{0.01\linewidth} \\
         \includegraphics[width=0.25\linewidth,height=2.5cm]{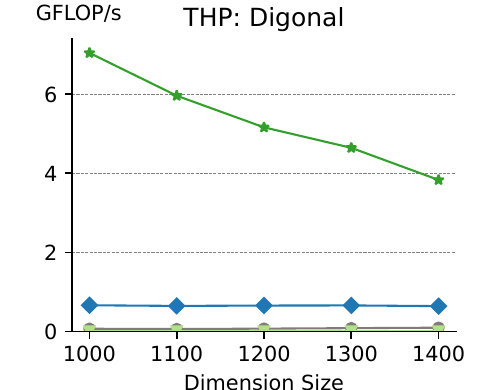} \hspace{0.045\linewidth} \vspace{0.01\linewidth} & \includegraphics[width=0.25\linewidth,height=2.5cm]{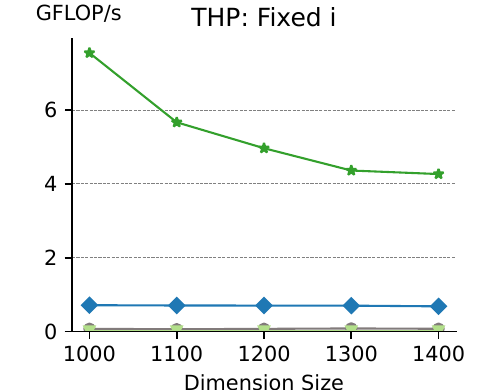} \hspace{0.045\linewidth} \vspace{0.01\linewidth} & \includegraphics[width=0.25\linewidth,height=2.5cm]{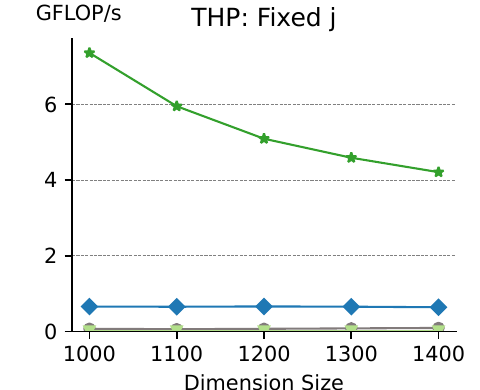}\vspace{0.01\linewidth} \\
         \includegraphics[width=0.25\linewidth,height=2.5cm]{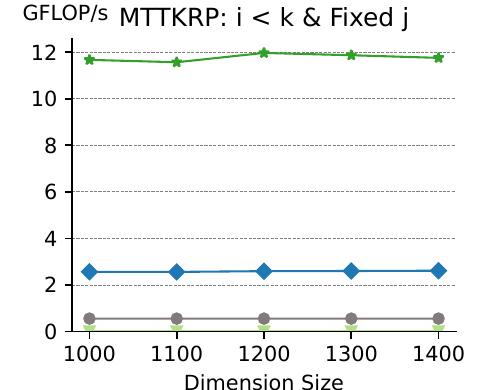} \hspace{0.045\linewidth} \vspace{0.01\linewidth} & \includegraphics[width=0.25\linewidth,height=2.5cm]{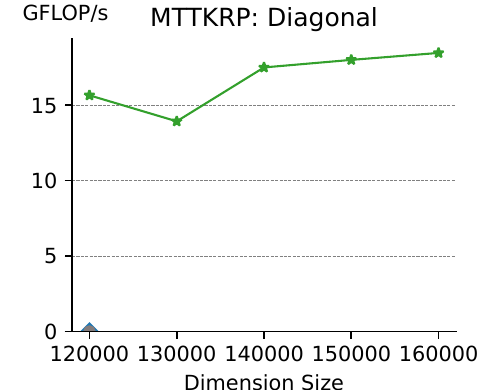} \hspace{0.045\linewidth} \vspace{0.01\linewidth} & 
         \includegraphics[width=0.25\linewidth,height=2.5cm]{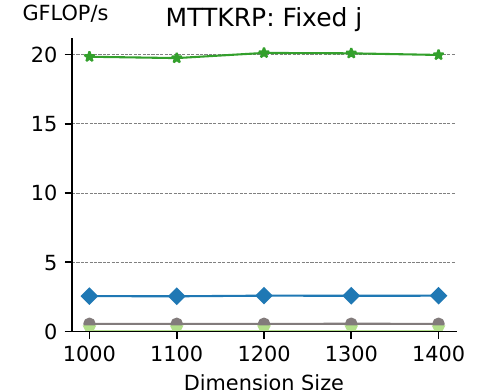}\vspace{0.01\linewidth} \\
         \includegraphics[width=0.25\linewidth,height=2.5cm]{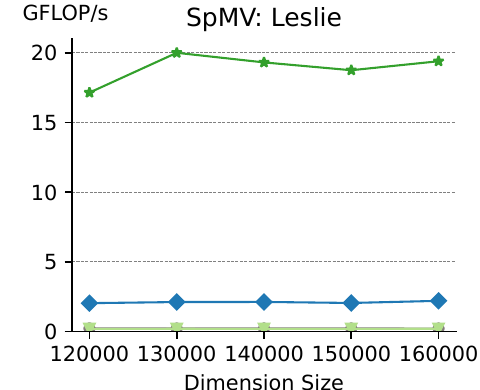} \hspace{0.045\linewidth} \vspace{0.01\linewidth} & \includegraphics[width=0.25\linewidth,height=2.5cm]{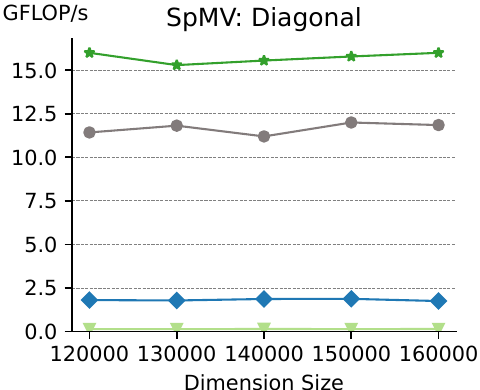} \hspace{0.045\linewidth} \vspace{0.01\linewidth} & \includegraphics[width=0.25\linewidth,height=2.5cm]{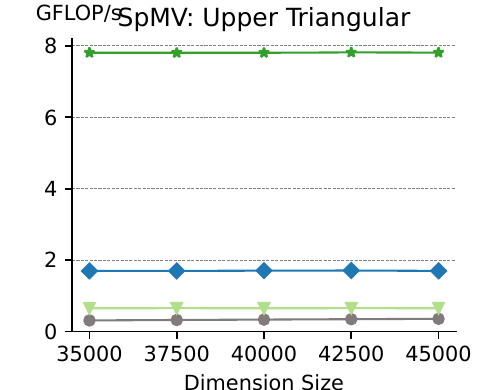} \\
         \includegraphics[width=0.25\linewidth,height=2.5cm]{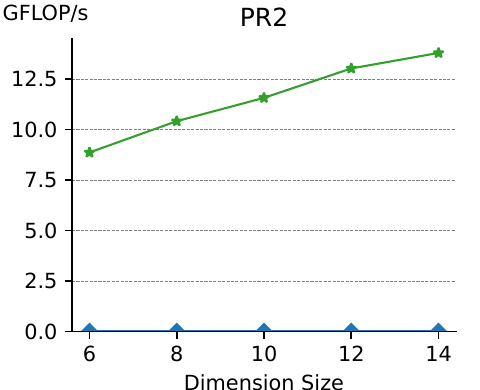} \hspace{0.045\linewidth} \vspace{0.01\linewidth} & \includegraphics[width=0.25\linewidth,height=2.5cm]{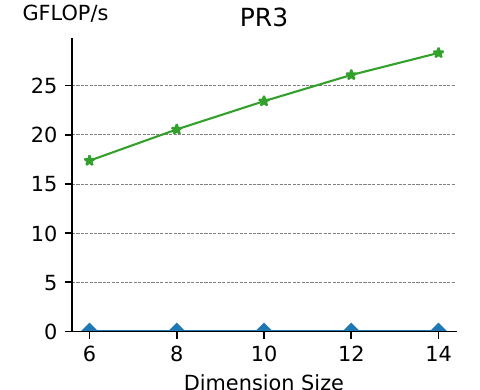} \hspace{0.045\linewidth} \vspace{0.01\linewidth} & \includegraphics[width=0.25\linewidth,height=2.5cm]{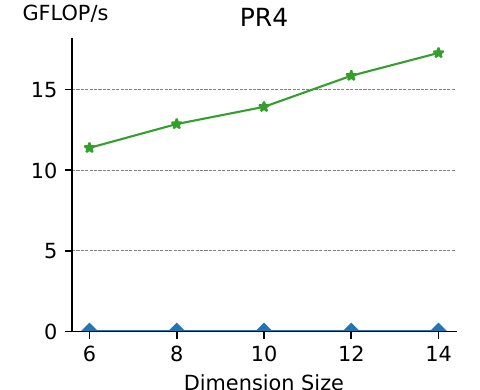}
    \end{tabular} 
    \vspace{-0.5cm}
    \caption{Performance comparison of various frameworks on ten threads on several structured tensor kernels.  In almost all cases, \systemname{} shows significantly better performance in comparison with all other competitors.}
    \label{fig:e2e:multithread}
    \vspace{-0.5cm}
\end{figure}

The number of floating point operations (FLOPs) is computed by multiplying the number of structure-aware iterations for each kernel by the summation of the number of additions and multiplications in that kernel. We used the structure-aware FLOPs since all the systems are leveraging the structure. We set a timeout of ten seconds for all frameworks. All the dense frameworks timed out on all kernels except SpMV with an upper triangular structure and half of TTM with an upper triangular structure. Therefore, we do not include them in any of the figures. The missing points for TACO are due to timeout (for polynomial regression kernels) or segmentation fault because of excessive memory allocation (other kernels). On MTT\_D, all competitors had a segmentation fault (excessive memory allocation) for sizes bigger than 120000.


\begin{figure}
\setlength\tabcolsep{.5pt}
    \centering
    \begin{tabular}{c}
         \includegraphics[width=0.99\linewidth]{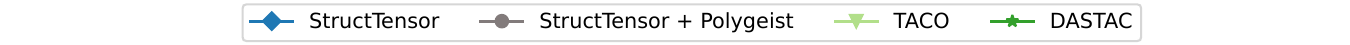}
    \end{tabular}
    \vspace{0.01\linewidth}
    \begin{tabular}{ccc}
         \includegraphics[width=0.25\linewidth,height=2.5cm]{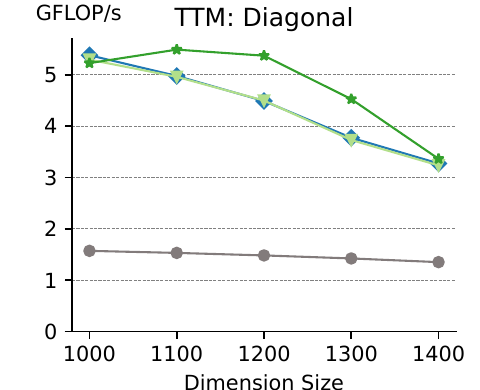} \hspace{0.045\linewidth} \vspace{0.01\linewidth} & \includegraphics[width=0.25\linewidth,height=2.5cm]{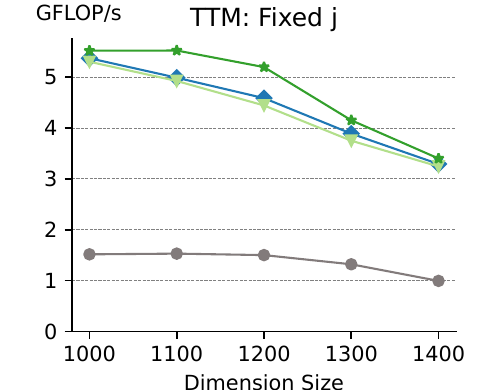} \hspace{0.045\linewidth} \vspace{0.01\linewidth} & \includegraphics[width=0.25\linewidth,height=2.5cm]{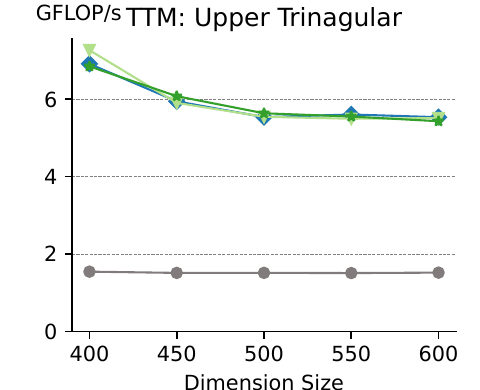}\vspace{0.01\linewidth} \\
         \includegraphics[width=0.25\linewidth,height=2.5cm]{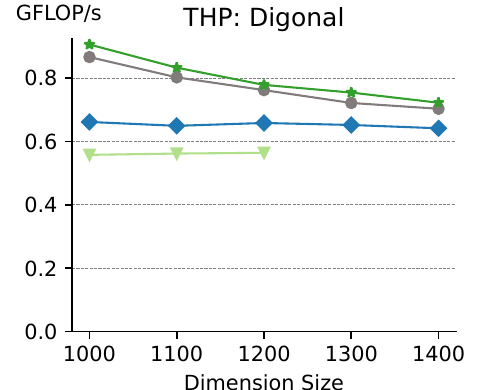} \hspace{0.045\linewidth} \vspace{0.01\linewidth} & \includegraphics[width=0.25\linewidth,height=2.5cm]{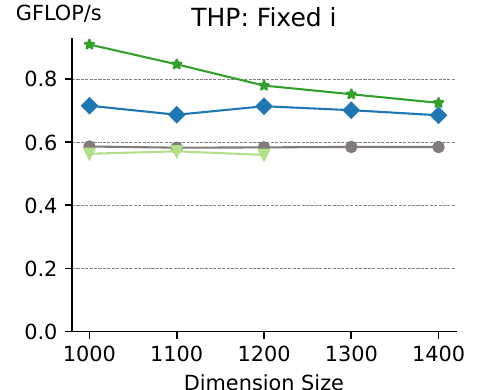} \hspace{0.045\linewidth} \vspace{0.01\linewidth} & \includegraphics[width=0.25\linewidth,height=2.5cm]{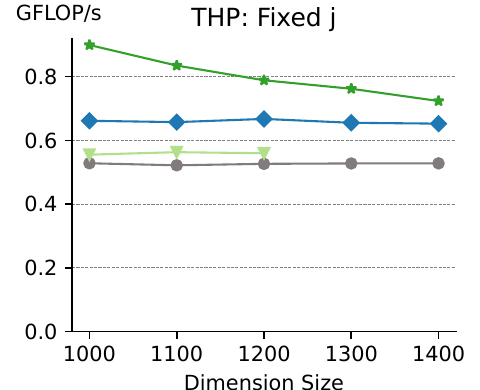}\vspace{0.01\linewidth} \\
         \includegraphics[width=0.25\linewidth,height=2.5cm]{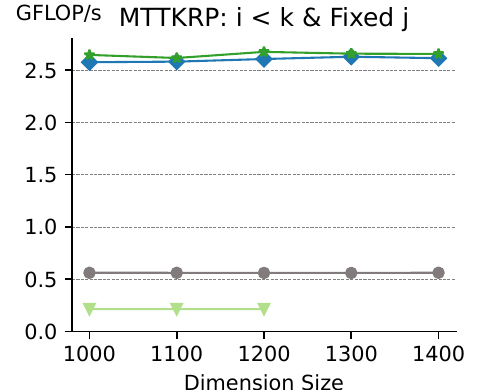} \hspace{0.045\linewidth} \vspace{0.01\linewidth} & \includegraphics[width=0.25\linewidth,height=2.5cm]{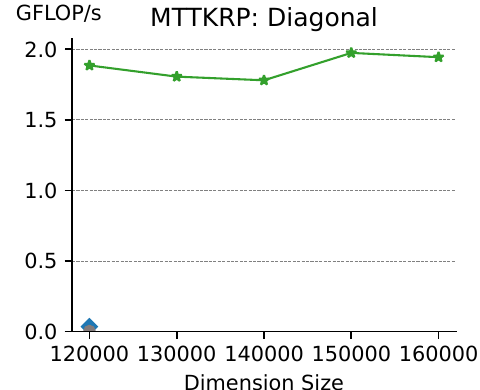} \hspace{0.045\linewidth} \vspace{0.01\linewidth} & 
         \includegraphics[width=0.25\linewidth,height=2.5cm]{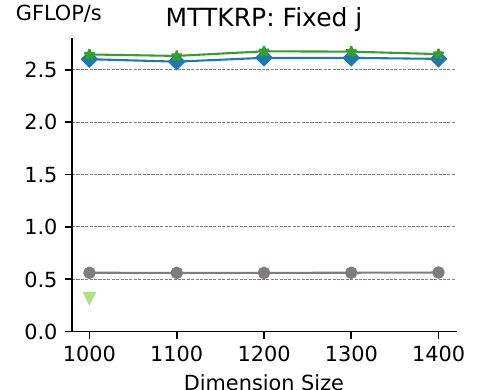}\vspace{0.01\linewidth} \\
         \includegraphics[width=0.25\linewidth,height=2.5cm]{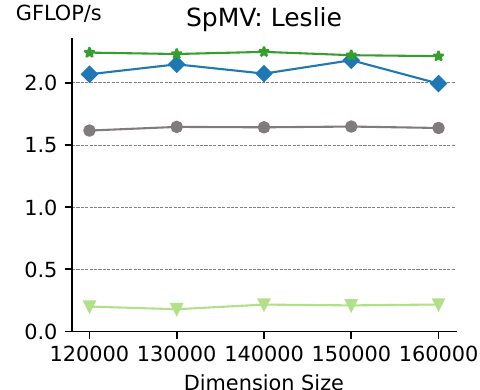} \hspace{0.045\linewidth} \vspace{0.01\linewidth} & \includegraphics[width=0.25\linewidth,height=2.5cm]{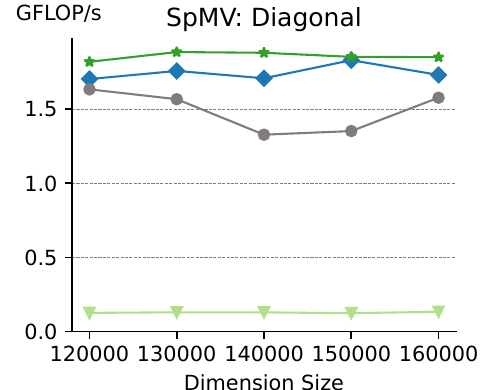} \hspace{0.045\linewidth} \vspace{0.01\linewidth} & \includegraphics[width=0.25\linewidth,height=2.5cm]{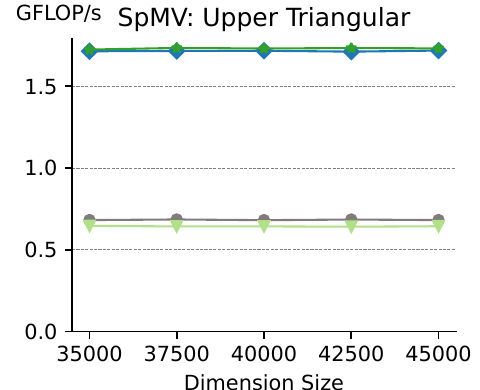} \\
         \includegraphics[width=0.25\linewidth,height=2.5cm]{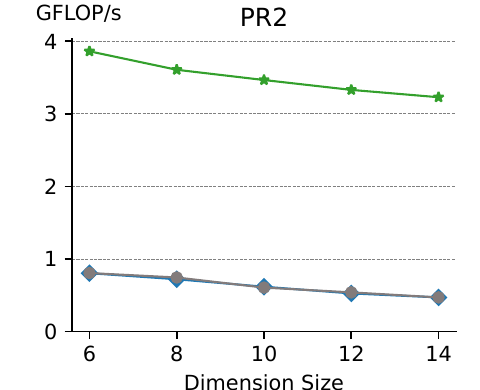} \hspace{0.045\linewidth} \vspace{0.01\linewidth} & \includegraphics[width=0.25\linewidth,height=2.5cm]{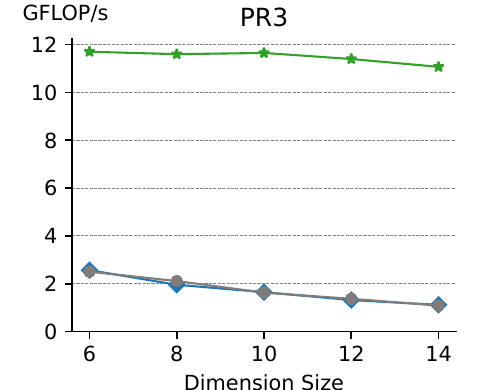} \hspace{0.045\linewidth} \vspace{0.01\linewidth} & \includegraphics[width=0.25\linewidth,height=2.5cm]{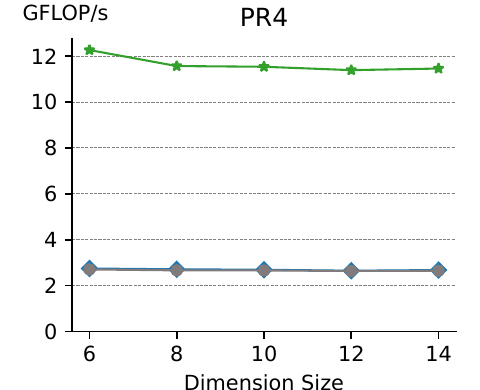}
    \end{tabular} 
    \vspace{-0.4cm}
    \caption{\Rev{Performance comparison} of various frameworks on one thread on several structured \Rev{tensor kernels}. In all cases, \systemname{} performs on par or better than \structtensor{}\Rev{ (w/ and w/o Polygeist).}}
    \label{fig:e2e:singlethread}
    \vspace{-0.4cm}
\end{figure}

\smartpara{Multi-Thread}
Figure~\ref{fig:e2e:multithread} compares the performance of each framework's best version. Static schedule is used for all the experiments. We only show the single-threaded performance results for systems that do not support parallelism (TACO\footnote{TACO parallelization problem is mentioned in this GitHub issue:
The link is removed due to anonymization.} for all kernels except SpMV and \structtensor{}). As shown in Figure~\ref{fig:e2e:multithread}, \systemname{} performs significantly better than all the competitors while running on multi threads. Overall, compared to the competitors, \systemname{} can get up to 50$\times$ performance enhancement. 

In all scenarios, \structtensor{} stands in second place from the performance perspective despite lacking parallelization. This shows the importance of capturing data structure and having an efficient algorithm that leverages this information. TACO's code cannot be parallelized on any of the kernels, except SpMV, which leads to slower performance. Polygeist automatic parallelization fails to parallelize the code in most cases. The reason behind it is discussed in Section~\ref{sec:experiment:parallel}. In the polynomial regression kernels, TACO times out, Polygeist parallelization descales, and \structtensor{} is running on a single thread; hence, their performance is significantly diminished. \systemname{} outperforms them by leveraging the flexibility and efficacy of sparse tensor algebra algorithms and merging them with optimizations and specializations provided by dense tensor algebra through polyhedral techniques. Combined with parallelization, these techniques lead to up to 1 to 2 orders of magnitude performance boost compared to structure-aware competitors. 

\smartpara{Single-Thread}
Figure~\ref{fig:e2e:singlethread} represents the performance of all frameworks running on a single thread. \systemname{} performs on par or outperforms all competitors in almost all cases while running on a single thread. TACO and \structtensor{} achieve closely to \systemname{} since they also leverage the structure. However, \systemname{} manages to outperform them in most cases (by up to 1 order of magnitude) by having a better memory layout, leading to better cache locality and vectorization. 


The downward slope on some kernels (e.g., TTM\_DP and TTM\_J) is due to increased cache miss rate (from 0.2\% to 12.6\% for TTM\_DP and from 0.2\% to 13.2\% for TTM\_J) by increasing the size (cache miss rate is measured through PAPI for all experiments, but not shown due to space constraints). The generated code by Polygeist fails to outperform \systemname{} despite leveraging the structure, manual data layout compression, and having out-of-the-box polyhedral and MLIR optimizations. This shows that using Polygeist to recover affine structure from a lower-level code and applying such optimizations out-of-the-box without a higher-level analysis does not benefit the performance.

\subsection{Memory Compression}
\label{sec:experiment:compress}

\begin{figure}
\setlength\tabcolsep{.5pt}
    \centering
    \begin{tabular}{c}
    \includegraphics[width=0.99\linewidth]{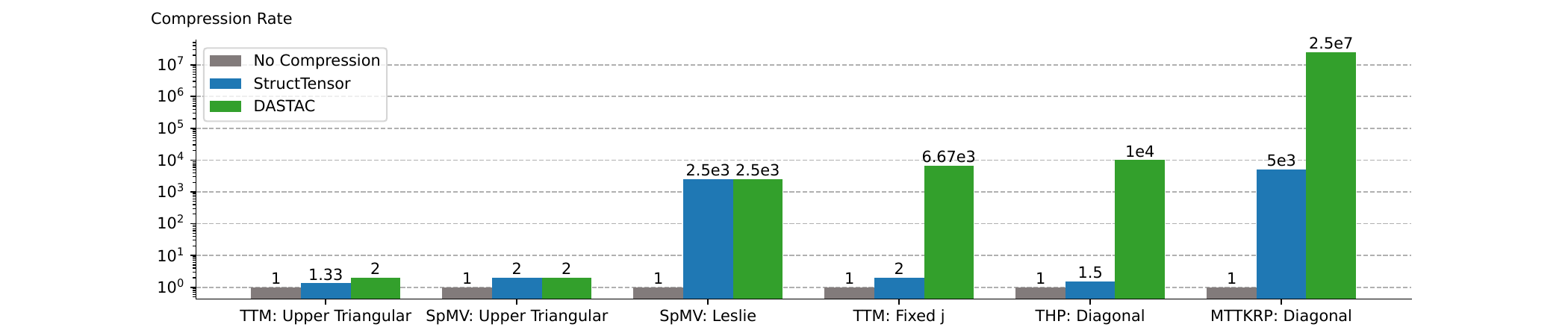}
    \end{tabular}
    \caption{Evaluation of compression impact on memory footprint of \systemname{} and \structtensor{} against an implementation \Rev{with no compression}. The dimension size for all kernels is $10000$. \systemname{} reaches the maximum compression on all tensors for each kernel, while \structtensor{} often does a sub-optimal job.}
    \label{fig:compression}
\end{figure}

\smartpara{Impact of Compression on Memory Footprint}
We evaluate the impact of the memory compression technique used by \systemname{} in comparison to an implementation without compression and \structtensor{}'s manual input compression on six different structures:  1) TTM\_UT, 2) SpMV\_UT, 3) SpMV\_L, 4) TTM\_J, 5) THP\_DP, and 6) MTT\_D (cf. Table~\ref{tbl:kernels}). We calculate the allocated memory for each structured kernel based on the number of elements allocated for all input and output tensors. The implementation \Rev{without compression} allocates memory for all the elements of the tensors. \structtensor{} reduces the allocated memory to unique elements of the input based on the structure and all output elements. \systemname{} has a better understanding of compression based on the compressed tensor computation and keeps only the elements required for tensor computation in both inputs and output. Therefore, even if a tensor is dense, but some of its elements are not used in the computation, \systemname{} compresses the tensor so that unused elements are not stored. 

All dimensions for all kernels are considered to be $10000$. The compression rate on the y-axis of Figure~\ref{fig:compression} is the memory allocated by the implementation \Rev{without compression} divided by the memory allocated by other systems. The higher compression rate leads to less memory complexity. Most elements are unique in kernels such as \Rev{TTM\_UT} and \Rev{SpMV\_UT}; therefore, the compression rate is low. Kernels such as \Rev{SpMV\_UT} and \Rev{SpMV\_L} only need the input tensors to be compressed since all the output elements are available and unique; hence, \structtensor{} and \systemname{} have the same compression rate. In all other cases, input and output tensors require compression in an ideal scenario. In several kernels, there are unique elements that are not used in the tensor computation \Rev{(e.g., MTT\_D)}. Since \systemname{} performs \Rev{computationally structure-aware} compression on these kernels, \systemname{}'s compression rate is significantly higher than \structtensor{}.


\begin{figure}
    \includegraphics[width=\linewidth]{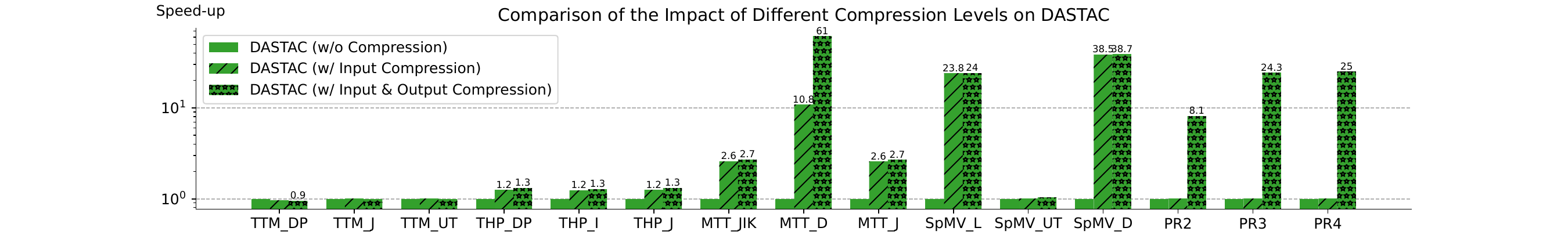}
    \caption{\Rev{Evaluation of the impact of different levels of compression on the performance of \systemname{}.}}
    \label{fig:compress_opt_dastac}
\end{figure}


\smartpara{Impact of Different Levels of Compression}
Figure~\ref{fig:compress_opt_dastac} shows that each level of compression improves the performance for the majority of kernels. Compressing the input has no impact on polynomial regression cases since the input is fully dense. In TTM cases, compression cannot improve the cache locality and the cache miss rate stays the same even after compression, hence, the performance is similar to uncompressed code.

In the tensor computations where the input compression significantly reduces the memory footprint, which leads to a significantly better cache miss rate, output compression has a minimal impact. On the other hand, in tensor expressions where the input is dense or compression over only the input cannot improve the cache accesses significantly, output compression has a larger impact on enhancing the performance. 

\begin{figure}
\setlength\tabcolsep{.5pt}
    \centering
    \begin{tabular}{ccccccccc}
        \includegraphics[width=0.12\linewidth]{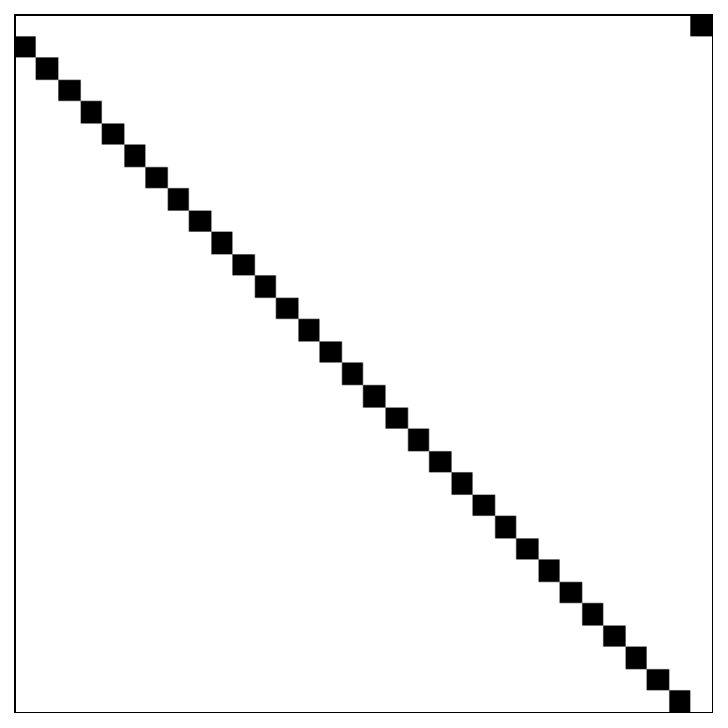} & 
        \includegraphics[width=0.02\linewidth]{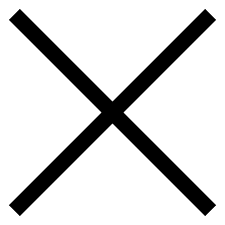} & 
        \includegraphics[width=0.12\linewidth]{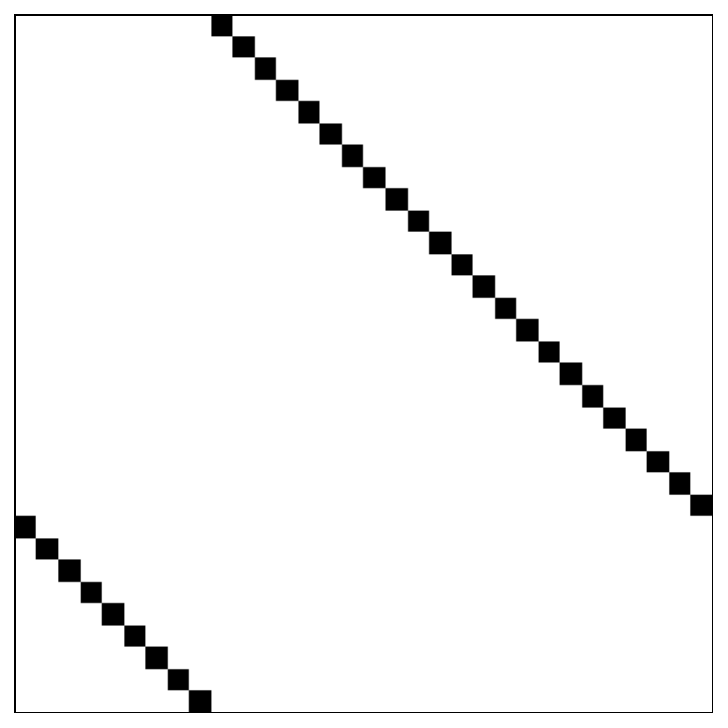} & 
        \includegraphics[width=0.05\linewidth]{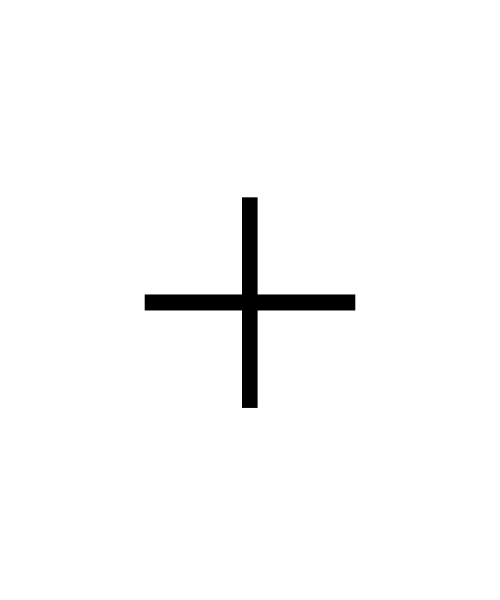} & 
        \includegraphics[width=0.12\linewidth]{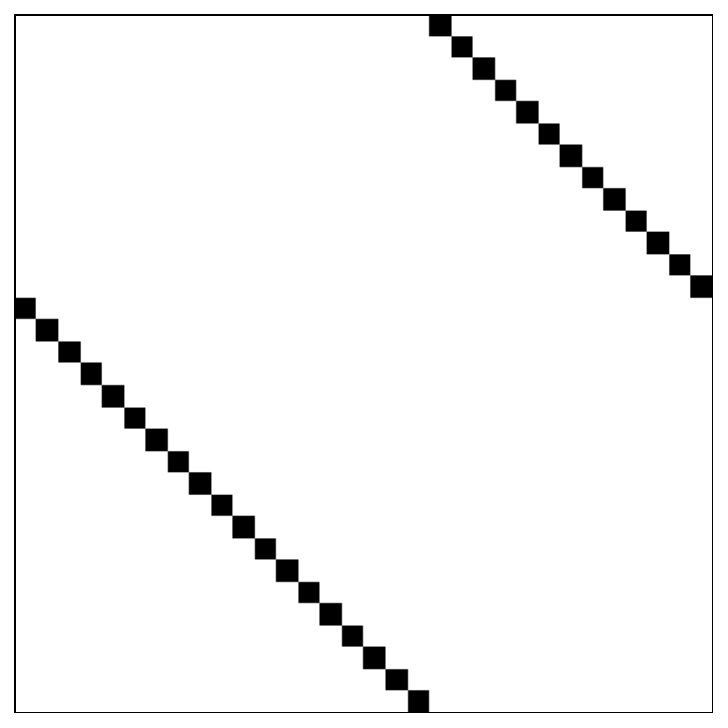} & 
        \includegraphics[width=0.02\linewidth]{figures/post-rebuttal/visualization/Times.png} & 
        \includegraphics[width=0.12\linewidth]{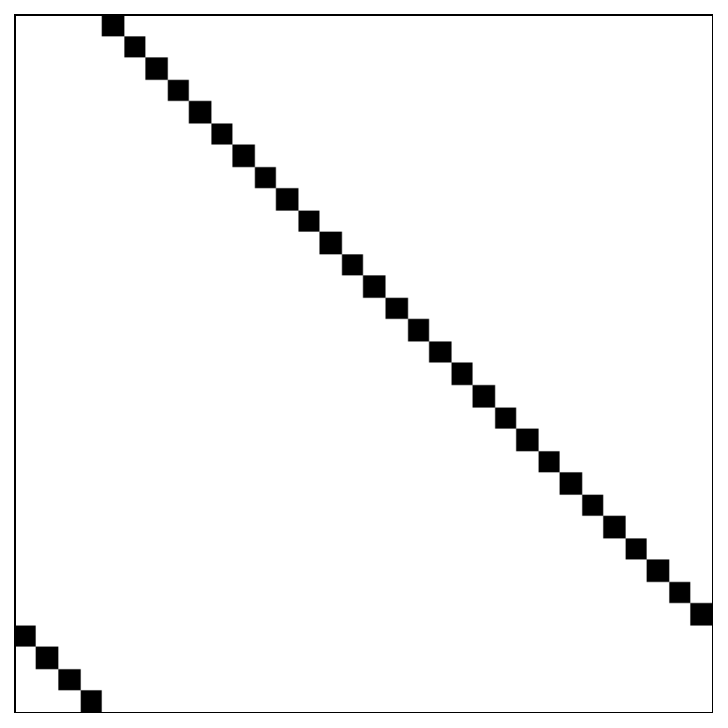} & 
        \includegraphics[width=0.05\linewidth]{figures/post-rebuttal/visualization/Plus.png} & 
        \dots
    \end{tabular}\\
    \begin{tabular}{ccc}
    \includegraphics[width=0.25\linewidth,height=2.5cm]{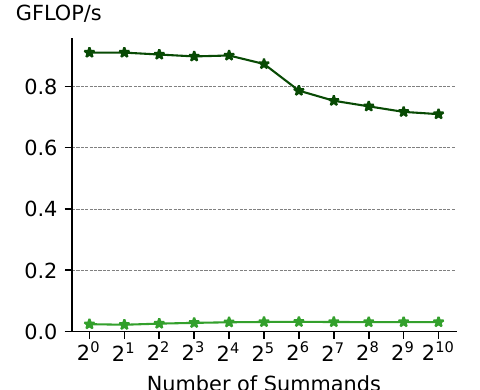} & 
    \includegraphics[width=0.25\linewidth,height=2.5cm]{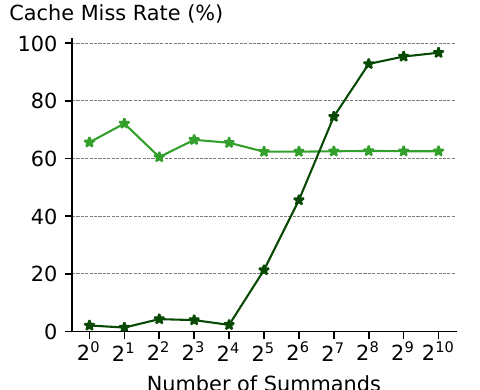} &
    \includegraphics[width=0.25\linewidth,height=2.5cm]{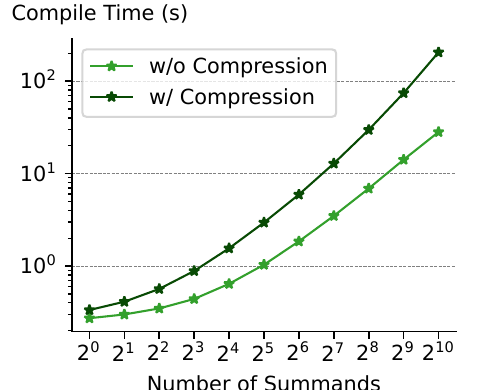}
    \end{tabular}
    \caption{\Rev{Evaluation of connection between compression impact and the number of summands using \systemname{}. A scaled version of the used structure for this experiment is shown on top.}}
    \label{fig:csd}
\end{figure}

\subsection{Impact of Structure's Complexity on Compression}
\label{sec:experiment:complex}

In this section, we introduce four new complex structures to study the impact of structure complexity on \systemname{}'s compression technique: 1) strided diagonal to study the impact of the number of summands in a tensor operation, 2) block butterfly factor matrix to study the impact of the number of the polyhedrals in the structure, 3) strided band to study the impact of distance of elements, and 4) sub-triangular to study the impact of density.




\smartpara{Number of Summands in Tensor Computation and Compression}
To measure the connection between these two factors (cf. Figure~\ref{fig:csd}), we consider the following tensor computation:

\begin{tabular}{rcl}
   $A(i, k)$  & $:=$ &  $\sum_{w=1}^{p} B^w(i, j) * C^w(i, k)$
\end{tabular}

\noindent
where $B$ and $C$ have a strided diagonal structure with a random stride. In other words:

\begin{tabular}{rcl}
   $B^w_U(i, j)$  & $:=$ &  $(0 \leq i < N) * (0 \leq j < N) * ((j - i) \% N = s_B^w)$
\end{tabular}

\noindent
The size $N$ is fixed to $2^{14}$ here, and the stride $s_B^w$ is selected randomly for each tensor. The output of each of these multiplications is a strided diagonal with stride $s_B^w + s_C^w$. We vary the number of summands from $1$ to $2^{10}$ in powers of $2$. More summands result in longer compilation time. Generating code for the compressed version takes longer than without compression since the compression indexing should be computed for more summands. For the compressed version, as the size increases, the number of elements fitting in the cache decreases since the compression cannot compensate anymore; therefore, the cache miss rate goes from almost $0$ to almost $100\%$. The performance starts dropping on the same size that the cache miss rate starts increasing. The generated code without compression has a consistent cache miss rate around $60\%$, and its performance is also consistently low. This is due to not caching because of high sparsity.

\begin{figure}
\setlength\tabcolsep{.5pt}
    \centering
    \begin{tabular}{ccccc}
         \includegraphics[width=0.12\linewidth]{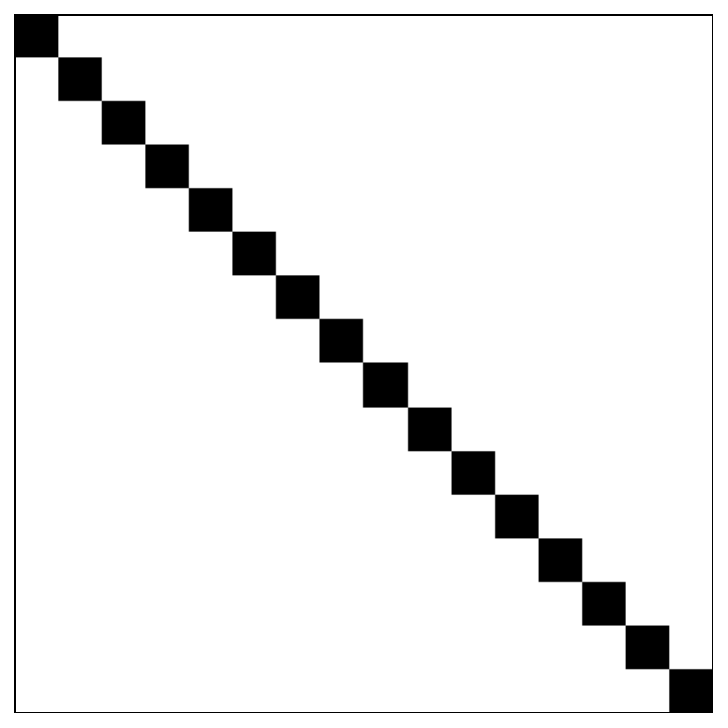} \hspace{0.05\linewidth} &
         \includegraphics[width=0.12\linewidth]{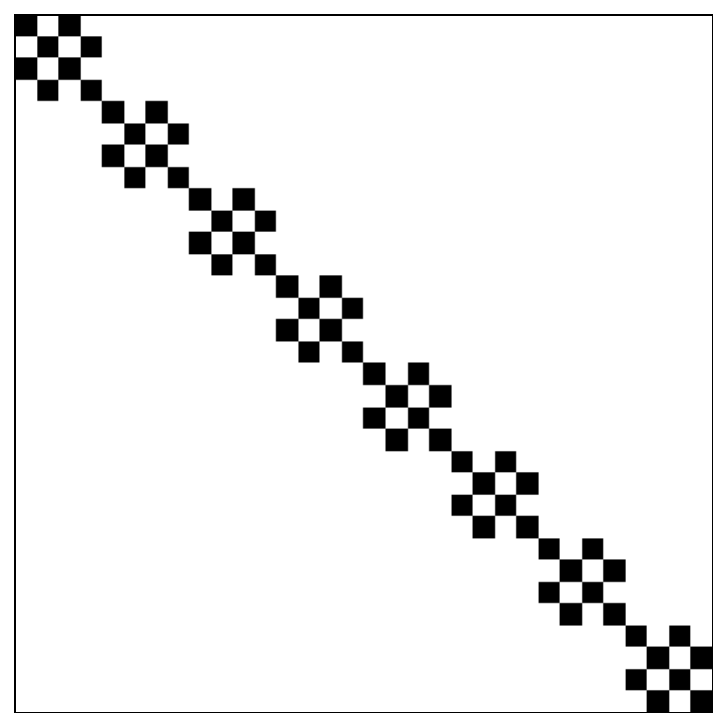} \hspace{0.05\linewidth} &
         \includegraphics[width=0.12\linewidth]{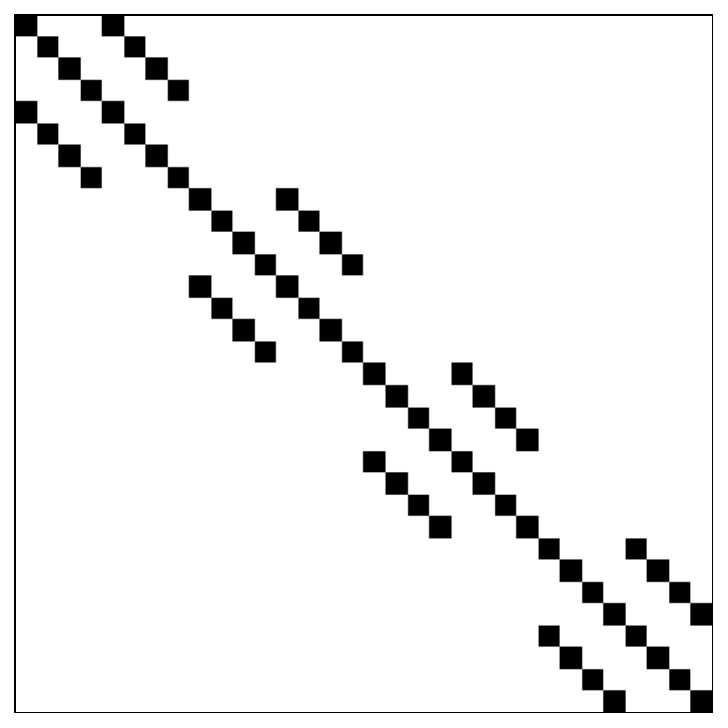} \hspace{0.05\linewidth} &
         \includegraphics[width=0.12\linewidth]{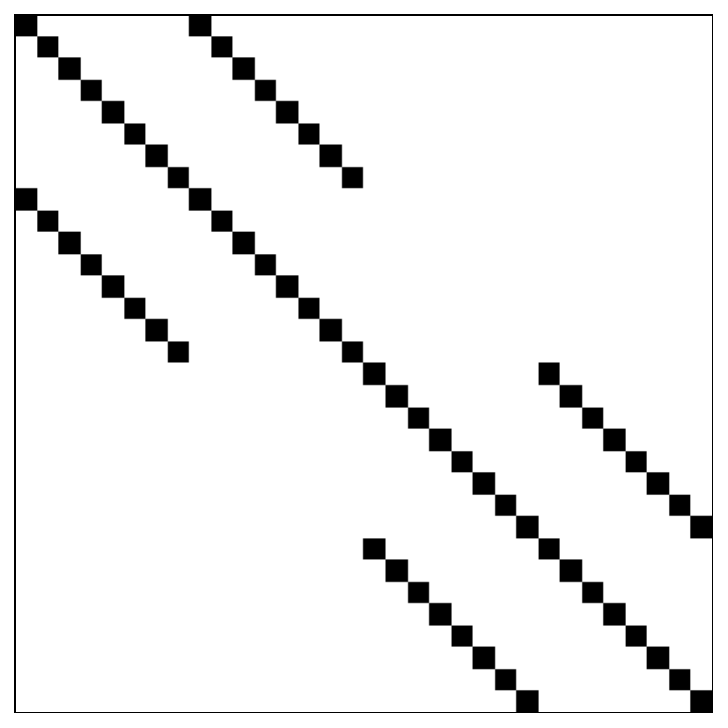} \hspace{0.05\linewidth} &
         \includegraphics[width=0.12\linewidth]{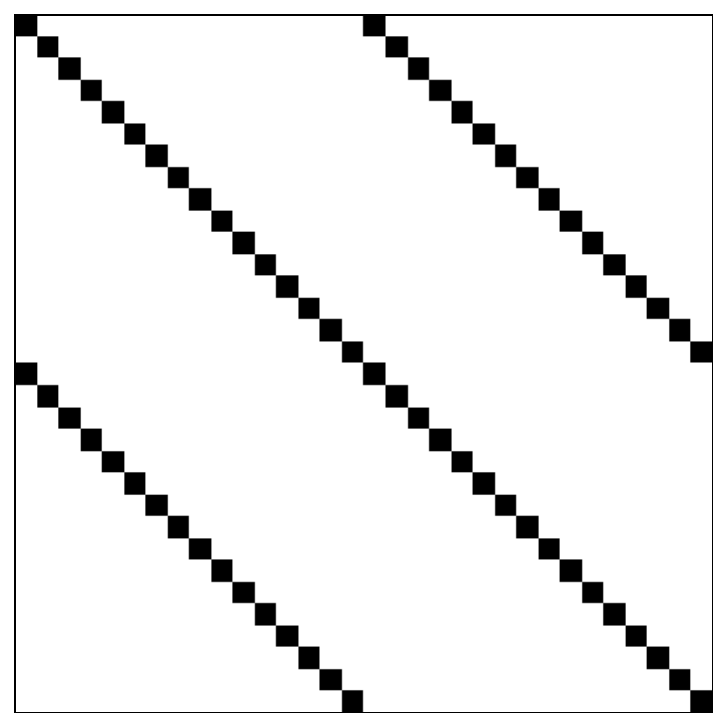}
    \end{tabular} \\
    \begin{tabular}{ccc}
    \includegraphics[width=0.25\linewidth,height=2.5cm]{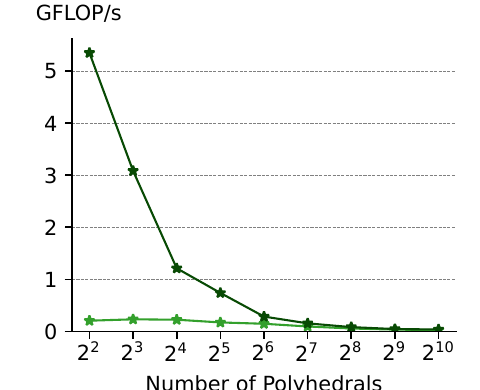} & 
    \includegraphics[width=0.25\linewidth,height=2.5cm]{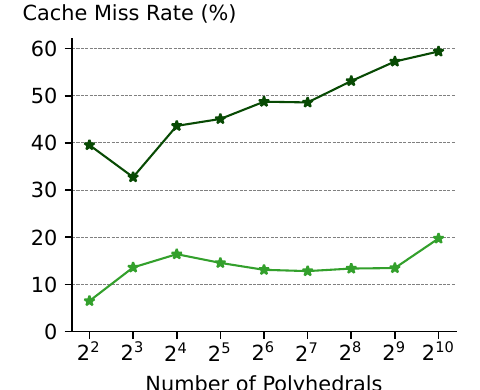} &
    \includegraphics[width=0.25\linewidth,height=2.5cm]{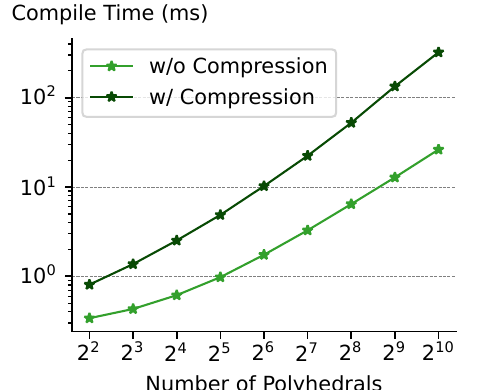}
    \end{tabular}
    \caption{\Rev{Evaluation of connection between compression impact and the number of polyhedrals using \systemname{}. A scaled version of the used structure for this experiment is shown on top.}}
    \label{fig:butterfly}
\end{figure}
\smartpara{Number of Polyhedrals and Compression}
To measure the connection between these two factors (cf. Figure~\ref{fig:butterfly}), we considered a SpMV computation where the matrix dimension is $2^{14}$ and has a block butterfly factor matrix~\cite{ChenDLY0RR22}. The structure is as follows:

\begin{tabular}{c}
   $B_U(x, y)$ $:=$ $(0 \leq i < b) * (0 \leq j < b) * (0 \leq s < k/2) * (0 \leq c < 2) * (0 \leq d < 2) *$ \\
   $(0 \leq r < \frac{n}{k}) * (x = r*k*b+(\frac{c*b*k}{2})+b*s+i) * (y = r*k*b+(\frac{d*b*k}{2})+b*s+j)$
\end{tabular}

\noindent
Matrix $B$ is an $nb \times nb$ matrix. We set $n=2^{10}$ and $b=2^{4}$. To vary the number of polyhedrals, we unroll this unique set over the variables $c$, $d$, and $s$ and write it as a summation/union format. We also change the value of $k$ from $2$ to $2^{10}$ in powers of two. This way, the matrix size and density stay exactly the same. However, the underlying structure and the number of polyhedrals change. Therefore, we can measure the impact of changing the number of polyhedrals alone without other factors interfering. The number of polyhedrals is $2k$. Increasing the number of polyhedrals increases the compilation time for compressed and uncompressed code generation with a similar slope. The cache miss rate for the compressed version increases by increasing the number of polyhedrals since the data is compressed in several buffers rather than fewer contiguous buffers; hence the performance drops. The cache miss rate and performance are consistent for the uncompressed version. The cache miss rate is lower than the compressed version because the iteration space already traverses $b \times b$ blocks, and compression disrupts the contiguous nature of the buffer.

\begin{figure}
\setlength\tabcolsep{.5pt}
    \centering
    \begin{tabular}{cccccc}
         \includegraphics[width=0.12\linewidth]{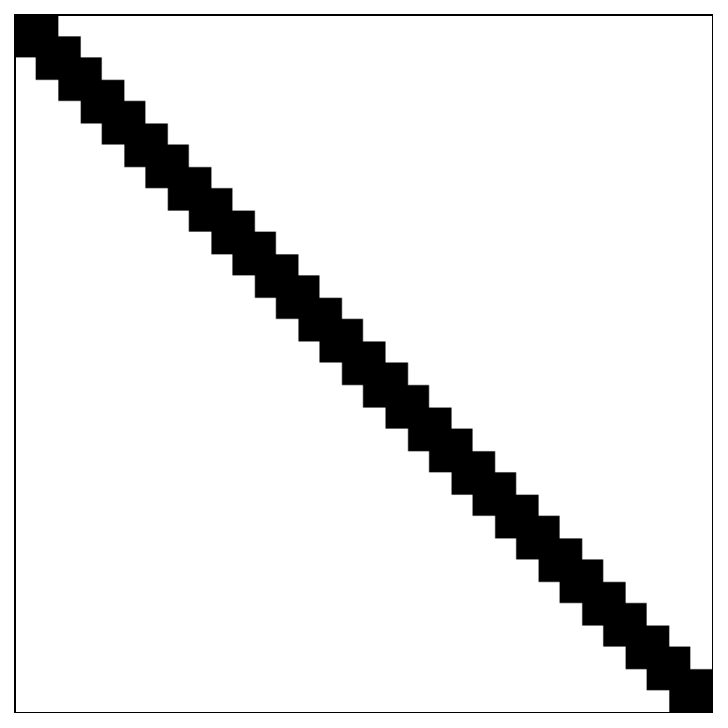} \hspace{0.03\linewidth} &
         \includegraphics[width=0.12\linewidth]{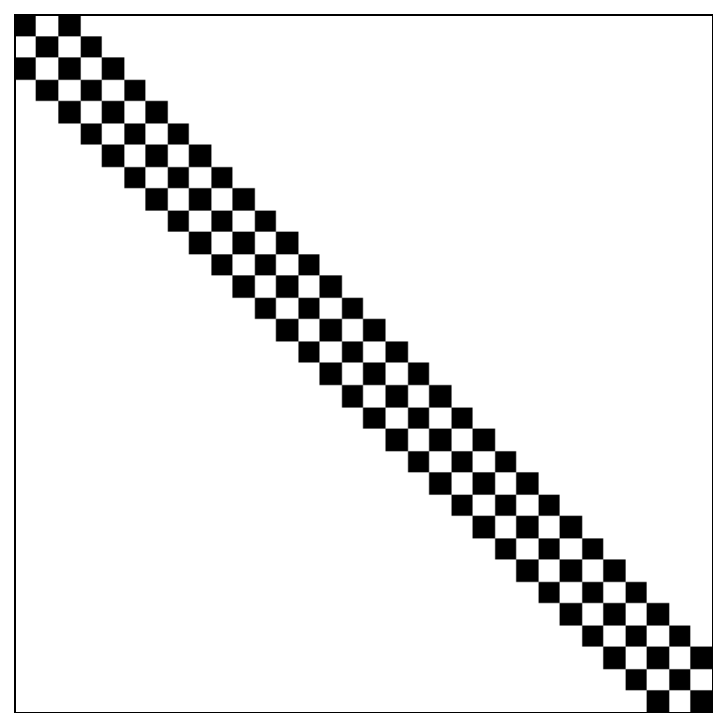} \hspace{0.03\linewidth} &
         \includegraphics[width=0.12\linewidth]{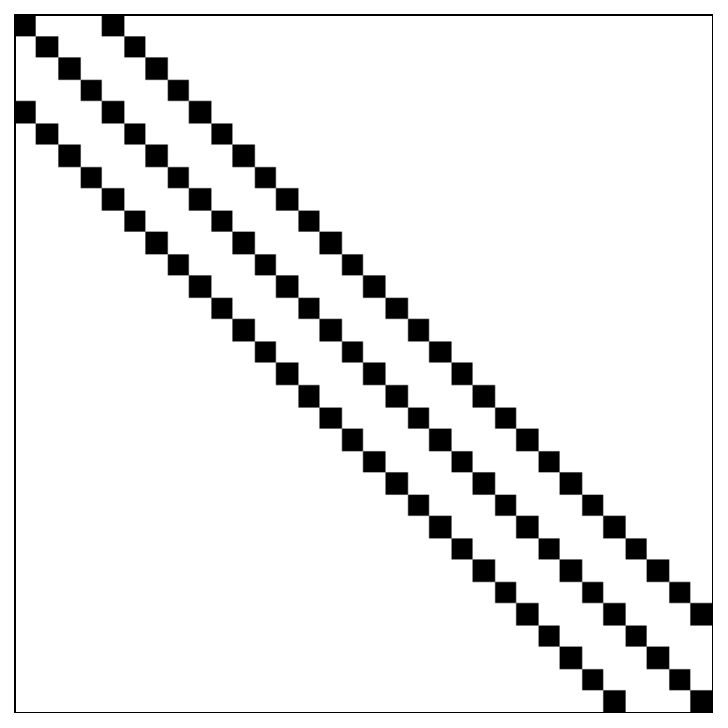} \hspace{0.03\linewidth} &
         \includegraphics[width=0.12\linewidth]{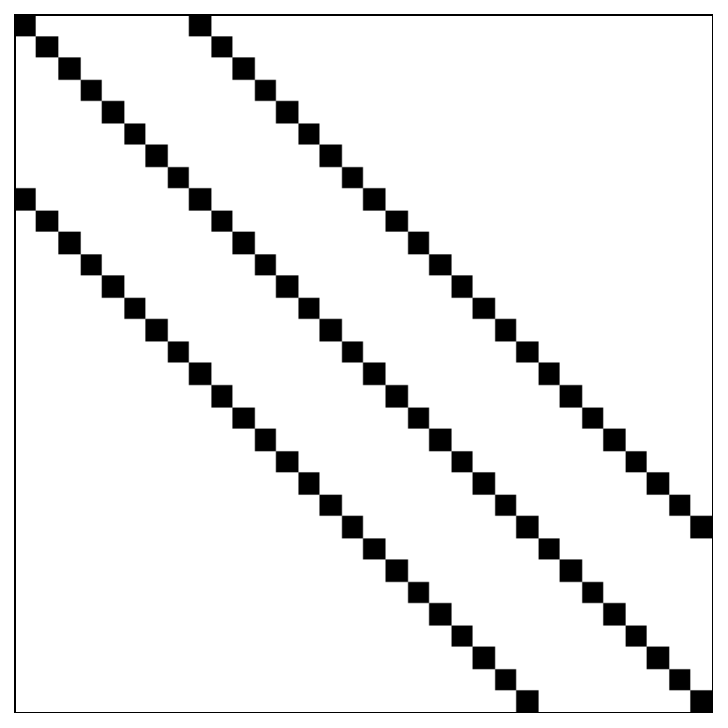} \hspace{0.03\linewidth} &
         \includegraphics[width=0.12\linewidth]{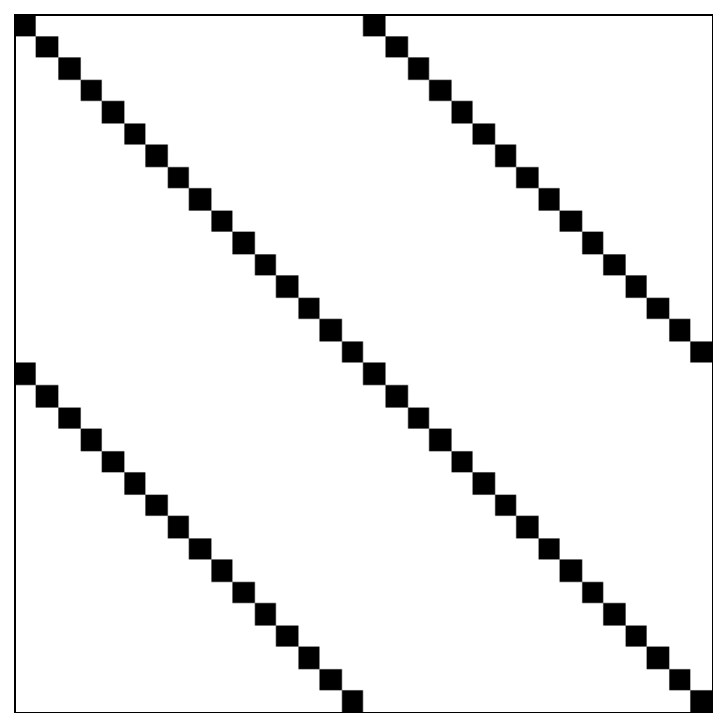} \hspace{0.03\linewidth} &
         \includegraphics[width=0.12\linewidth]{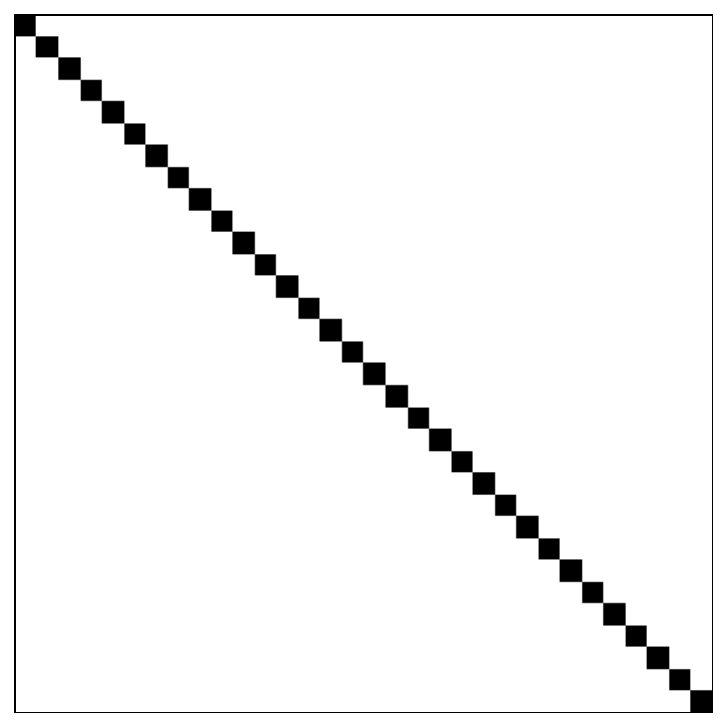}
    \end{tabular}
    \begin{tabular}{cc}
    \includegraphics[width=0.4\linewidth]{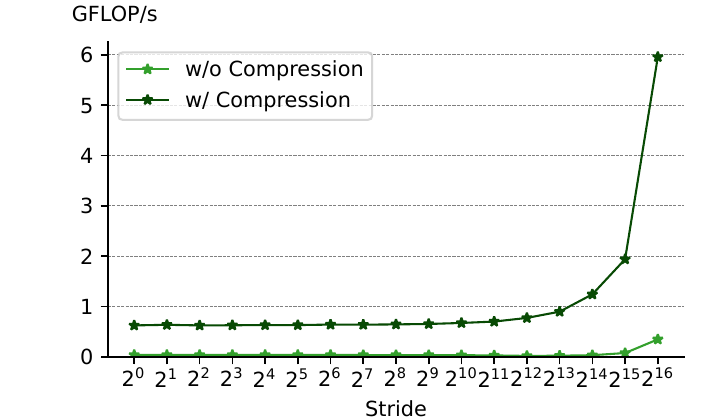} & 
    \includegraphics[width=0.4\linewidth]{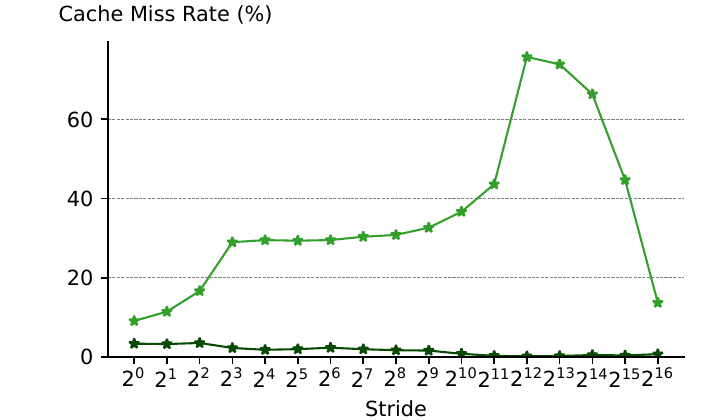}
    \end{tabular}
    \caption{\Rev{Evaluation of connection between compression impact and stride of the matrix using \systemname{}. A scaled version of the used structure for this experiment is shown on top.}}
    \label{fig:stride}
\end{figure}

\smartpara{Distance of Elements and Compression}
To measure the connection between these two factors (cf. Figure~\ref{fig:stride}), we considered a SpMV computation where the matrix dimension is $2^{16}$ and has a strided band matrix structure. The structure is as follows:

\begin{tabular}{rcl}
   $B_U(i, j)$  & $:=$ &  $(0 \leq i < N) * (j = i) + (0 \leq i < N) * (j - i = s) + (0 \leq i < N) * (i - j = s)$
\end{tabular}

\noindent
We vary the stride $s$ from $1$ (tridiagonal) to $2^{16}$ (diagonal). By increasing $s$, the number of elements is reduced but their distance is increased. The compressed version mostly fits in the cache and is only impacted by the number of elements. The fewer the elements, the higher the performance and the lower the cache miss rate for the compressed version. The uncompressed version's cache miss rate increases as the distance of the elements increases. However, when the distance reaches beyond $2^{12}$, the impact of the decrease in the number of elements helps the cache miss rate to drop. Both versions reach maximum performance when the matrix is diagonal.

\begin{figure}
\setlength\tabcolsep{.5pt}
    \centering
    \begin{tabular}{cccccc}
         \includegraphics[width=0.12\linewidth]{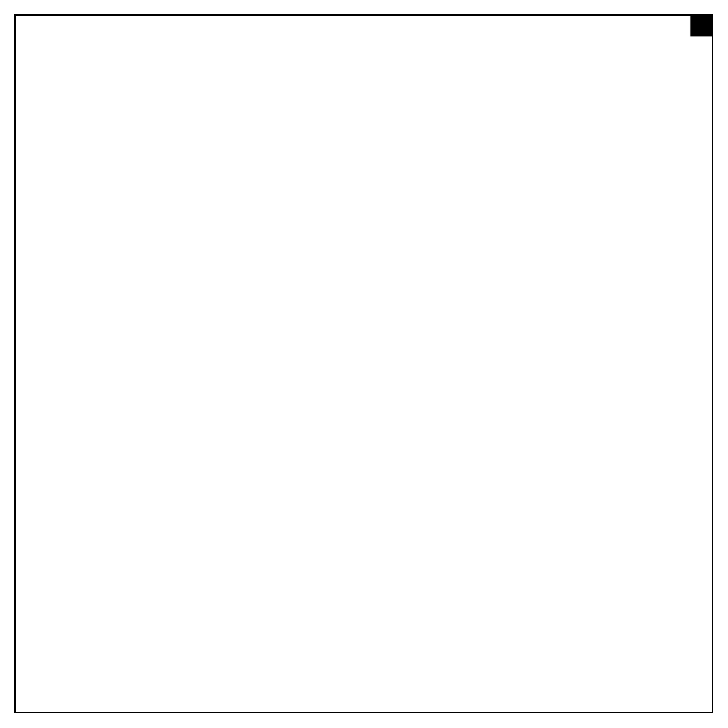} \hspace{0.03\linewidth} &
         \includegraphics[width=0.12\linewidth]{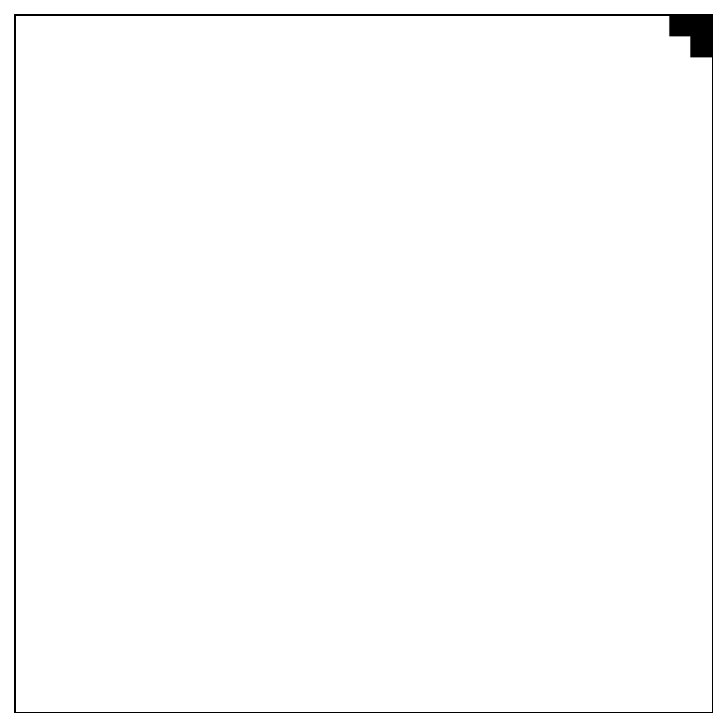} \hspace{0.03\linewidth} &
         \includegraphics[width=0.12\linewidth]{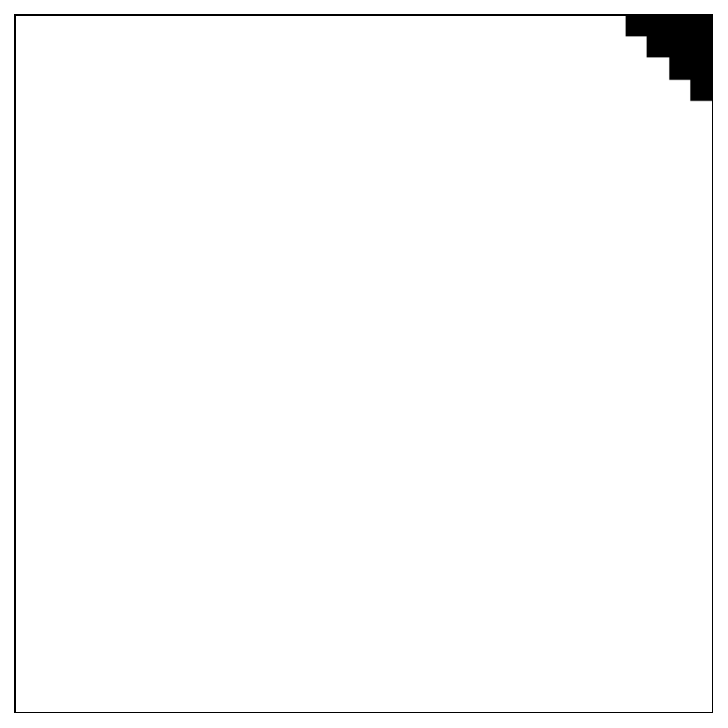} \hspace{0.03\linewidth} &
         \includegraphics[width=0.12\linewidth]{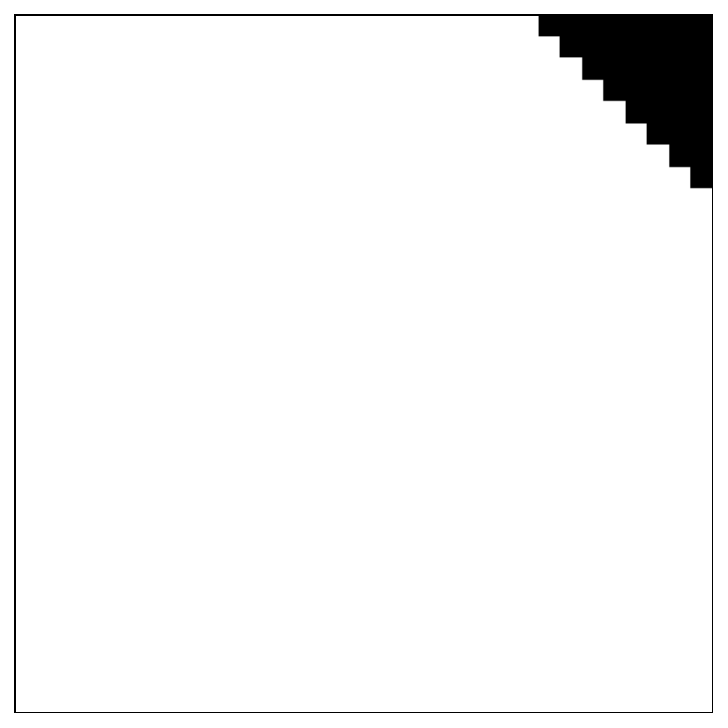} \hspace{0.03\linewidth} &
         \includegraphics[width=0.12\linewidth]{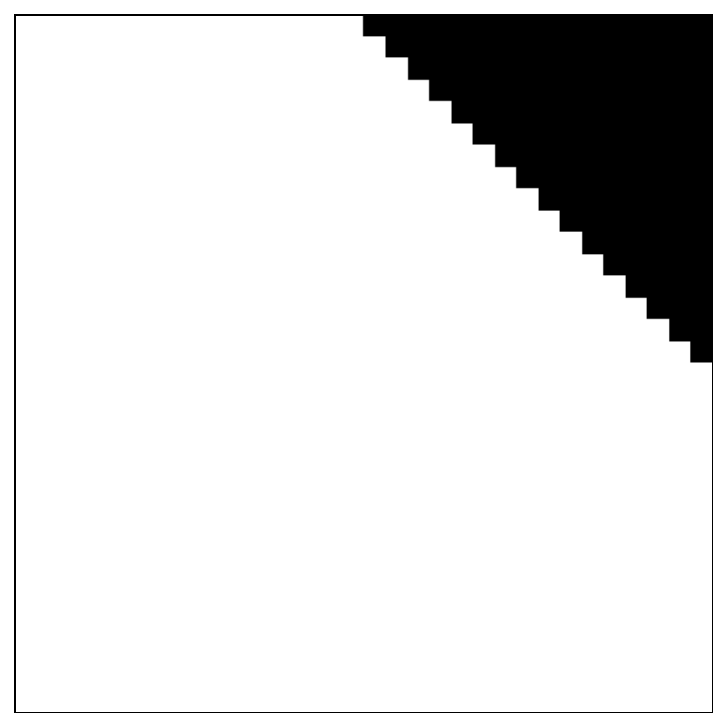} \hspace{0.03\linewidth} &
         \includegraphics[width=0.12\linewidth]{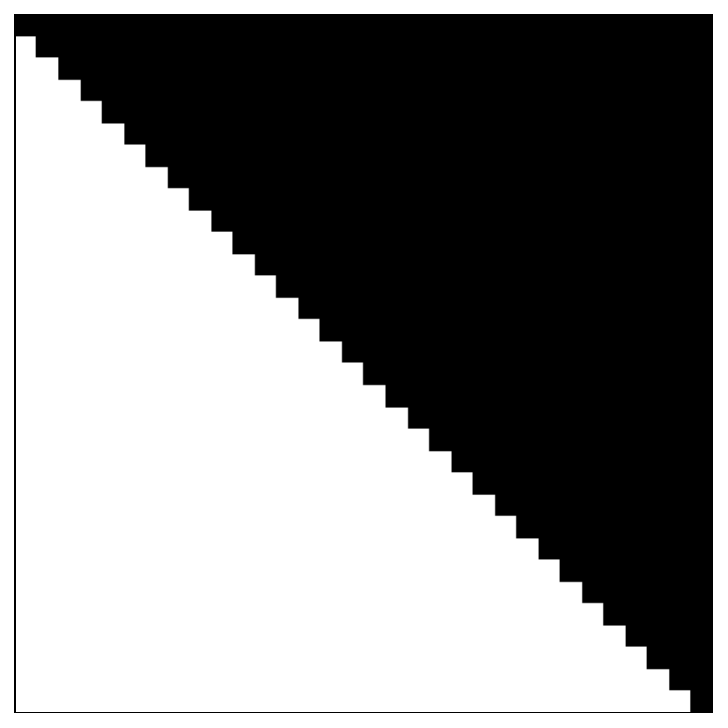}
    \end{tabular}
    \begin{tabular}{cc}
    \includegraphics[width=0.4\linewidth]{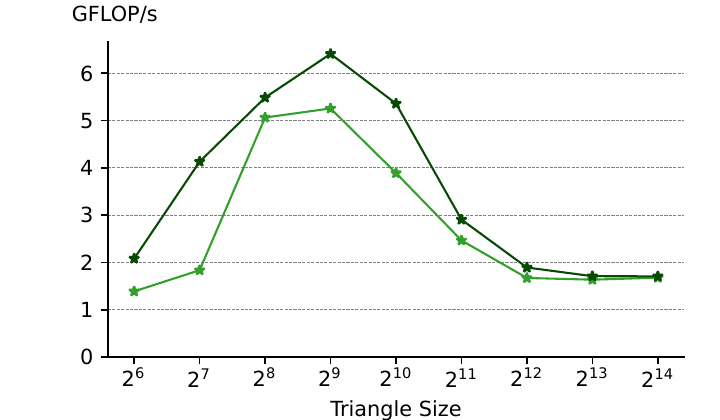} & 
    \includegraphics[width=0.4\linewidth]{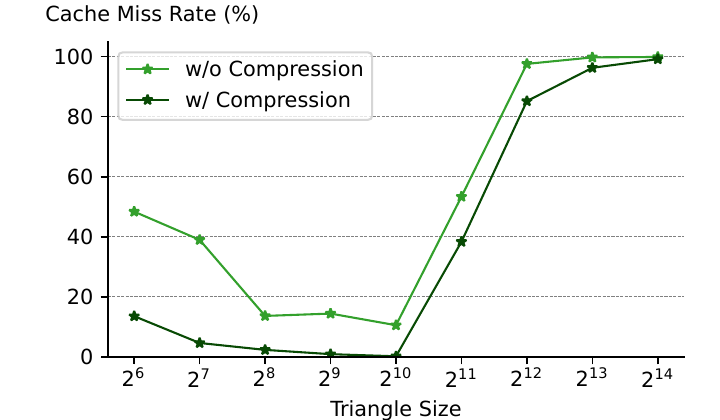}
    \end{tabular}
    \caption{\Rev{Evaluation of connection between compression impact and the density of the matrix using \systemname{}. A scaled version of the used structure for this experiment is shown on top.}}
    \label{fig:density}
\end{figure}


\begin{figure}
\setlength\tabcolsep{.5pt}
    \centering
    \begin{tabular}{c}
         \includegraphics[width=0.99\linewidth]{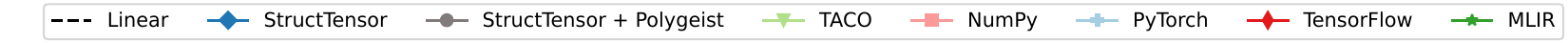}
    \end{tabular}
    \vspace{0.01\linewidth}
    \begin{tabular}{ccc}
         \includegraphics[width=0.25\linewidth,height=2.5cm]{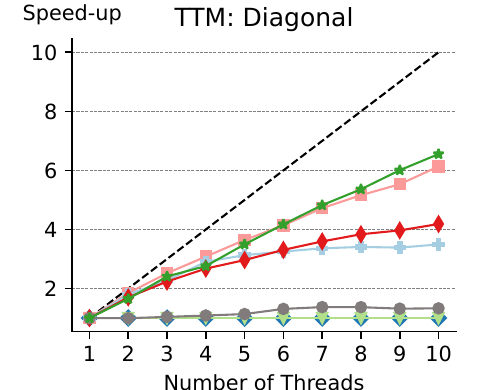} \hspace{0.045\linewidth} \vspace{0.01\linewidth} & \includegraphics[width=0.25\linewidth,height=2.5cm]{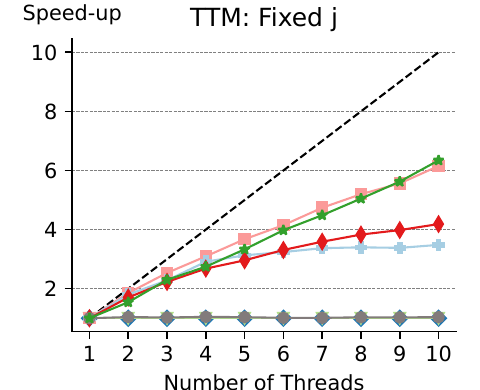} \hspace{0.045\linewidth} \vspace{0.01\linewidth} & \includegraphics[width=0.25\linewidth,height=2.5cm]{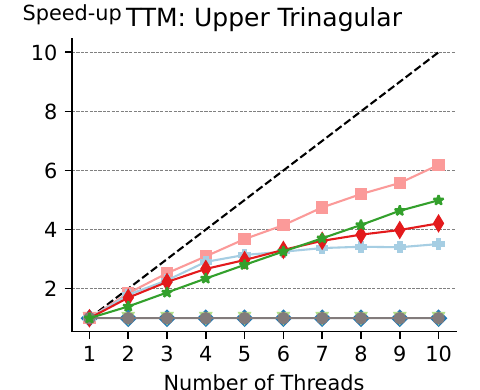}\vspace{0.01\linewidth} \\
         \includegraphics[width=0.25\linewidth,height=2.5cm]{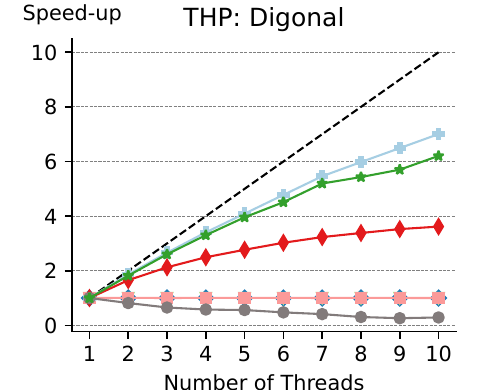} \hspace{0.045\linewidth} \vspace{0.01\linewidth} & \includegraphics[width=0.25\linewidth,height=2.5cm]{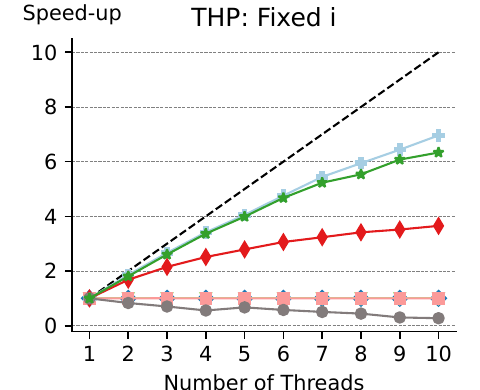} \hspace{0.045\linewidth} \vspace{0.01\linewidth} & \includegraphics[width=0.25\linewidth,height=2.5cm]{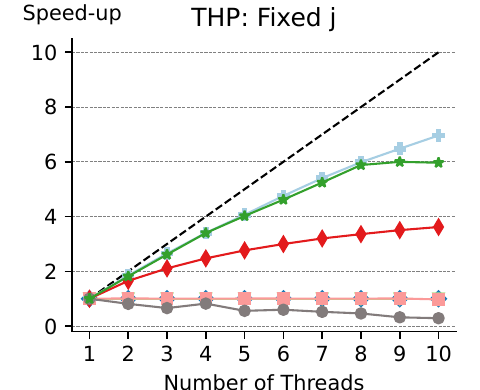}\vspace{0.01\linewidth} \\
         \includegraphics[width=0.25\linewidth,height=2.5cm]{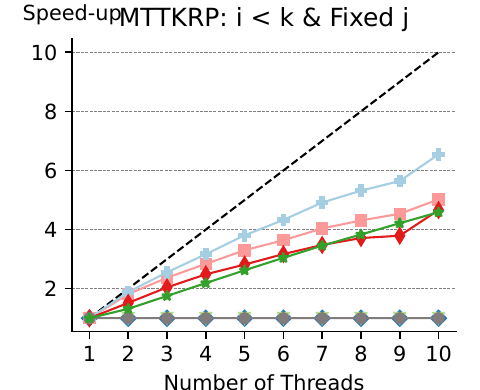} \hspace{0.045\linewidth} \vspace{0.01\linewidth} & \includegraphics[width=0.25\linewidth,height=2.5cm]{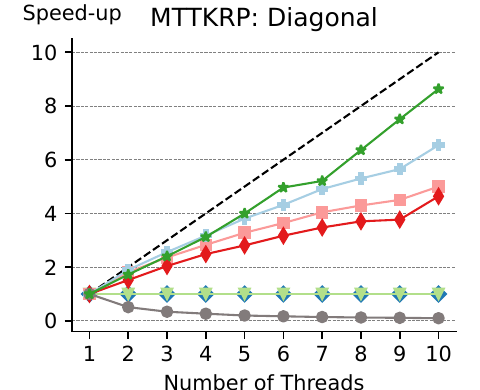} \hspace{0.045\linewidth} \vspace{0.01\linewidth} & 
         \includegraphics[width=0.25\linewidth,height=2.5cm]{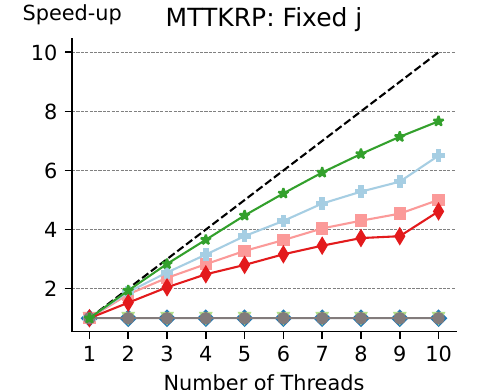}\vspace{0.01\linewidth} \\
         \includegraphics[width=0.25\linewidth,height=2.5cm]{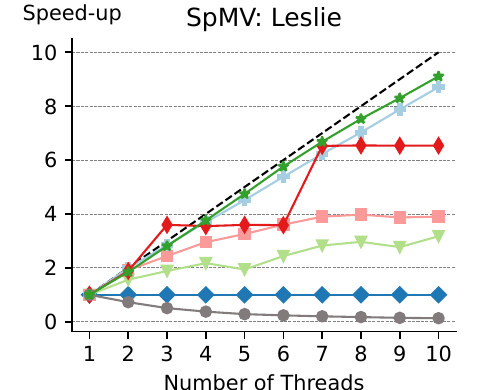} \hspace{0.045\linewidth} \vspace{0.01\linewidth} & \includegraphics[width=0.25\linewidth,height=2.5cm]{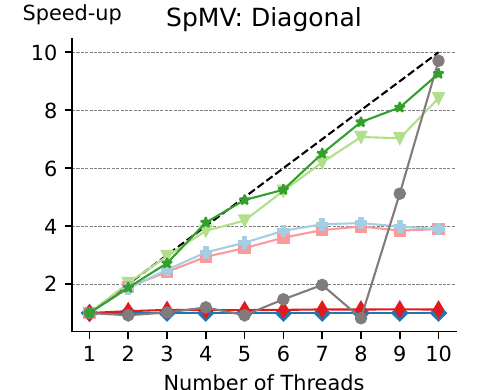} \hspace{0.045\linewidth} \vspace{0.01\linewidth} & \includegraphics[width=0.25\linewidth,height=2.5cm]{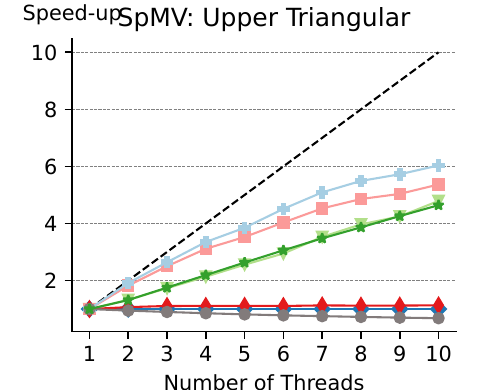} \\
         \includegraphics[width=0.25\linewidth,height=2.5cm]{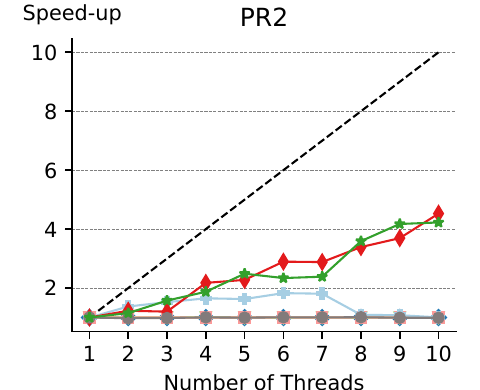} \hspace{0.045\linewidth} \vspace{0.01\linewidth} & \includegraphics[width=0.25\linewidth,height=2.5cm]{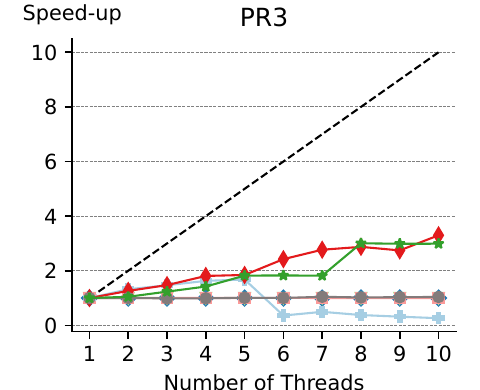} \hspace{0.045\linewidth} \vspace{0.01\linewidth} & \includegraphics[width=0.25\linewidth,height=2.5cm]{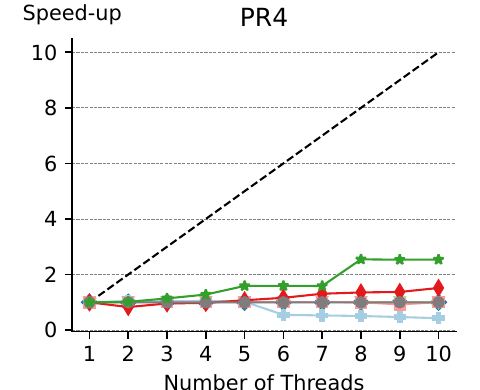}
    \end{tabular} 
    \vspace{-0.4cm}
    \caption{Comparison of the speed-up scaling of various frameworks on different numbers of threads on several structured tensor operation kernels.  In almost all cases, \systemname{} scales on par or better than the state-of-the-art tensor algebra frameworks.}
    \label{fig:scale}
\end{figure}

\smartpara{Density and Compression}
To measure the connection between these two factors (cf. Figure~\ref{fig:density}), we considered a SpMV computation where the matrix dimension is $2^{14}$ and has a sub-triangular structure, meaning that the shape is triangular but only contains a subset of a upper triangular matrix. The structure is as follows:

\begin{tabular}{rcl}
   $B_U(i, j)$  & $:=$ &  $(0 \leq i < N) * (0 \leq j < N) * (j - i \geq N - k)$
\end{tabular}

\noindent
Here, $k$ is the size of the triangle. We vary $k$ from $2^6$ to $2^{14}$ in powers of two. Up to $k=2^9$, the density is low enough to keep all the elements in the cache for the compressed version. Both cases have their best performance on $k=2^9$. After that point, the density and number of elements become larger, hence the compression technique starts losing its benefit compared to uncompressed code. They finally merge into the same cache miss rate and the same performance when the matrix is upper triangular ($k=2^{14}$).

\subsection{Parallelization Scalability}
\label{sec:experiment:parallel}

We evaluate the \systemname{}'s ability to scale on all kernels mentioned in Table~\ref{tbl:kernels} by varying the number of threads from one to ten (cf. 
 Figure~\ref{fig:scale}). \Rev{A considerably large size is selected for all kernels to ensure that the frameworks do not timeout or generate errors (e.g., segmentation fault)}. The scaling factor on the y-axis shows the speed-up of each framework compared to running them sequentially.
As shown in Figure~\ref{fig:scale}, \systemname{} can scale almost linearly in most cases by increasing the number of threads. In nearly all cases, \systemname{} scales on par or better than state-of-the-art tensor algebra frameworks. In \Rev{several THP, MTTKRP, and SpMV} kernels, \systemname{} reaches an almost linear speed-up compared to running it on one thread. TACO \Rev{only parallelizes on SpMV kernels. \structtensor{} and All other kernels of TACO} can not be run in parallel; therefore, they do not scale by varying the number of threads. Only \systemname{} can consistently scale up on all kernels. \Rev{PyTorch fails to scale on polynomial regression kernels,} NumPy fails to scale on THP kernels, and TensorFlow fails to scale on \Rev{SpMV\_UT} and \Rev{SpMV\_D}. \Rev{Polygeist does not scale or even de-scales on the majority of kernels. The automatic parallelization provided by Polygeist leads to parallelizing inner loops and not the outer loops on a majority of kernels. This increases overhead while running on more threads, which leads to a decrease in performance. Furthermore, the code generated by Polygeist for SpMV\_D kernel shows an abnormal behavior on the lower number of threads, which appeared on a range of sizes.}

\subsection{Polygeist Performance Using the Structure}
\label{sec:experiment:opt}


\begin{figure}
    \includegraphics[width=\linewidth]{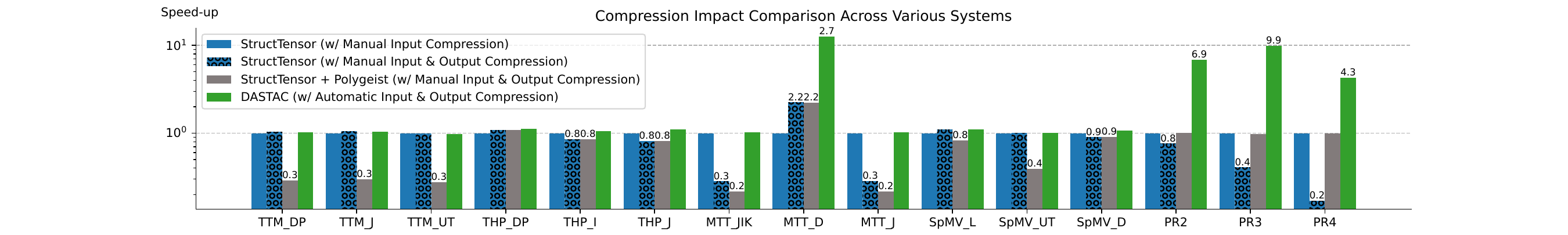} 
    \caption{\Rev{Evaluation of the impact of different systems' compression techniques on the performance.}}
    \label{fig:compress_opt_frameworks}
\end{figure}

\begin{figure}
\setlength\tabcolsep{.5pt}
    \centering
    \begin{tabular}{c}
    \includegraphics[width=0.99\linewidth]{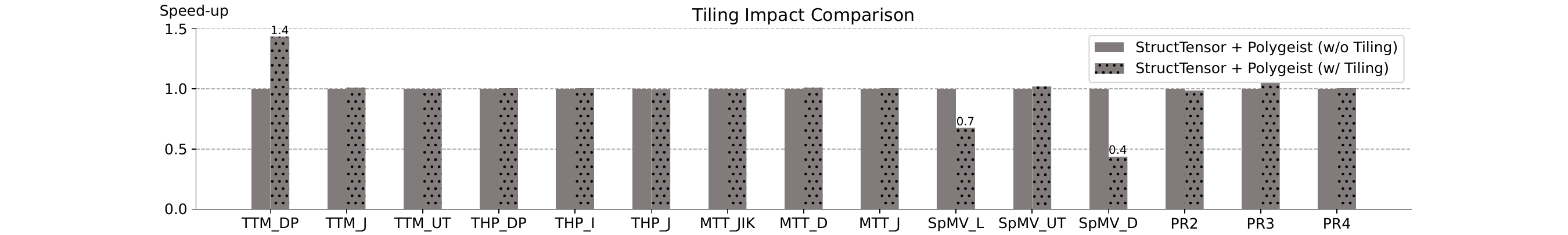} 
    \end{tabular}
    \caption{\Rev{Evaluation of the impact of Polygeist automatic tiling on the performance.}}
    \label{fig:tiling_opt}
\end{figure}

\Rev{Figure~\ref{fig:compress_opt_frameworks} shows that manually compressing both input and output almost always improves the performance, except for the cases with a hyper triangular structure, where the compression damages locality. Converting the manually compressed code to Polygeist is usually not helpful and in several cases, it causes performance degradation by removing the existing optimization opportunities for Clang. \systemname{} automatically compresses both input and output based on the iteration space and the compression algorithm almost always enhances the performance over all other competitors.}

\Rev{\smartpara{Impact of Automatic Tiling Using Polygeist}}
\Rev{Figure~\ref{fig:tiling_opt} shows that out-of-the-box tiling provided by Polygeist does not enhance the performance. Polygeist relies on MLIR compile pass to tile the generated code. However, the out-of-the-box tiling has no impact on the majority of the cases. In cases such as SpMV\_D and SpMV\_L, the iteration space is one-dimensional, and tiling decreases the performance.}

\section{Related Work}
\label{sec:related}

In this section, we elaborate on the research that has been done on dense and sparse (unstructured and structured) tensor algebra, as well as data layout compression. 

\smartpara{Dense Tensor Algebra}
Tensor Comprehension~\cite{TensorComprehension} provides optimized code for CUDA kernels by designing a polyhedral-based JIT compiler. TensorFlow~\cite{tensorflow} and PyTorch~\cite{pytorch} are deep learning frameworks providing efficient CPU and GPU code for tensor operation kernels. NumPy~\cite{numpy} is a fundamental tool for numerical computing, providing optimized code for dense tensor algebra computations. TVM~\cite{tvm} proposes a compiler to address machine learning-specific challenges, making it portable to various hardware devices. All of these systems generate significantly efficient code for dense tensor algebra computations. However, they lack leveraging the structure of the underlying data. isl~\cite{isl} and CLooG~\cite{cloog} are generic polyhedral-based frameworks that provide advanced loop optimizations and scheduling and are capable of efficient affine code generation, but they are not full-fledged compiler frameworks and thus lack the ability to densely pack the data layout. 

\smartpara{Unstructured Sparse Tensor Algebra}
The sparse polyhedral framework~\cite{strout2018sparse} combines the support for sparse tensor algebra with polyhedral techniques. TACO~\cite{kjolstad:2017:taco,DBLP:journals/pacmpl/ChouKA18} proposes a generic way to handle computation over sparse tensor algebra. SDQL~\cite{DBLP:journals/pacmpl/ShaikhhaHSO22,schleich2023optimizing,DBLP:conf/cgo/ShaikhhaHH24} uses a functional language to compile sparse tensor programs. JAX~\cite{jax2018github} is a high-performance computing tool relying on the MLIR~\cite{mlir} infrastructure with preliminary support for sparse tensor algebra. SciPy~\cite{virtanen2020scipy} is an algebraic library supporting a variety of kernels supporting both sparse and dense tensor algebra. A DSL to capture sparsity structure for recursive, pointer-based data structures and perform optimized computation on them is provided in \cite{chou2022compilation}. Recent work on Register Tiling~\cite{register_tiling} relies on low-level optimizations to better exploit compute resources in those cases, but focuses on SpMM only. Those frameworks capture dynamic sparsity nicely. However, they spend extra memory storage to store indexing and go through indirect accesses for computation. \systemname{} automatically compresses the tensor data and uses the most efficient data layout. Moreover, \systemname{} has direct access to tensor elements by relying on the symbolic polynomial index rather than reading it from an auxiliary array.

\Rev{Given the interdependencies between the optimal data layout and the target architecture details, recent work has focused on composability~\cite{sparsetir} and interoperability~\cite{mosaic} of such transformations to autotune an optimal scheme. These techniques are complementary to our proposed compression technique; exposing symbolic indexing as a composable transformation and augmenting search spaces with its usage is worth exploring in further work.}

\smartpara{Structured Tensor Algebra}
LGen~\cite{spampinato2016basic} proposes a polyhedral-based algorithm to capture structure on small-scale fixed-size linear algebra computations and generate faster code. EGGS~\cite{DBLP:journals/cgf/TangSKPLP20} unrolls the computation tree to specialize the computation over the sparse data. Sympiler~\cite{sympiler} performs compile-time symbolic analysis to produce optimized code for a set of linear algebra kernels. A method to capture the unstructured sparsity of matrices in a sparse way to generate polyhedral-based efficient code for the SpMV kernel is proposed in~\cite{augustine2019generating}. All the mentioned works are specialized in structured linear algebra \Rev{and leverage a limited set of operations}. However, they are not usable for higher-order tensors and lack a proper automatic data layout compression, \Rev{and thus are not suitable competitors}. \structtensor{}~\cite{structtensor} captures the structure of tensor algebra computation and propagates it throughout the computation, leading to efficient C++ code generation. However, \structtensor{}'s support for data layout is limited to inputs and requires the user to provide it manually. \structtensor{} does not support an automatic data layout compression despite inferring the structure. Moreover, \structtensor{} does not leverage specialized dense tensor algebra computation and relies on the underlying compiler for optimizations, even though it lowers the problem of structured tensor algebra computation to dense tensor algebra computation. Finch~\cite{DBLP:journals/corr/abs-2404-16730,DBLP:conf/cgo/AhrensDKA23} supports unstructured and structured sparsity but shares similar limitations. \systemname{} supports structured computation over higher-order tensors. Furthermore, \systemname{} packs the structured data densely using a symbolic indexing algorithm and enables leveraging polyhedral- and affine-based optimizations.

\smartpara{Data Layout Compression}
Tiramisu~\cite{baghdadi2019tiramisu} proposes a polyhedral compiler producing efficient and portable sparse tensor algebra code by introducing several optimizations, including a hardware-specific data layout transformation on both CPU and GPU. Polyhedral-based techniques to infer a data layout transformation with a better locality suitable for stencil computation over dense tensor algebra are provided in~\cite{henretty2013stencil}.  PolyMage~\cite{polyblocks} utilizes polyhedral techniques to transform the data layout of intermediate computation results and generates affine code for it. All the aforementioned frameworks propose and utilize advanced data layout transformation techniques for enhancing the data locality most suitably for the underlying hardware or based on the computation graph. However, they are only permuting the data without compressing based on the structure of the data. \systemname{} automatically infers a data transformation to densely pack the sparse data and provides a compressed data layout for the input, intermediate, and output data. A method using the Barvinok counting algorithm~\cite{barvinok} to index iterations in a parallel loop nest to collapse them for coarser parallelism through OpenMP is proposed in~\cite{clauss2017automatic}. However, this approach does not compress the data and is much more costly compared to \systemname{} since it computes the inverse of a quasi-polynomial.

\section{Conclusion}
In this paper, we introduce \systemname{}, the first code generation framework combining algorithmic optimizations known from sparse tensor algebra frameworks with performance-optimized low-level code known from decades of tuning dense tensors. 
The central design behind \systemname{} is a novel symbolic indexing algorithm to compress the structured tensor data into a contiguous memory buffer. 
This algorithm relies on the well-established mathematical foundation of polyhedral models, enabling additional optimizations such as \Rev{optimal conditional placements} and parallelization. Our experimental results show that the compressed data layout by \systemname{} achieves state-of-the-art memory consumption and sequential/parallel execution run time. In the future, we plan to target architectures such as GPU with applications such as sparse neural networks that can benefit from our proposed algorithm for densely packing tensor data. We also plan to consider other structured data including graphs. 

\section*{Acknowledgement}
This work was supported by the Engineering and Physical Sciences Research Council (EPSRC) grant EP/W007940/1 and partly funded by a gift from RelationalAI. The authors thank Huawei for their support of the distributed data management and processing laboratory at the University of Edinburgh.

\bibliography{main}

\end{document}